\documentclass[preprint, 3p,
authoryear]{elsarticle} %review=doublespace preprint=single 5p=2 column
\usepackage[hyphens]{url}

  \journal{Spatial and Spatio-temporal Epidemiology} % Sets Journal name

\usepackage{graphicx}
\usepackage[T1]{fontenc}
\usepackage{lmodern}
\usepackage{amssymb,amsmath}
\usepackage{lineno} % add
\usepackage{ifxetex,ifluatex}
\IfFileExists{upquote.sty}{\usepackage{upquote}}{}
\ifnum 0\ifxetex 1\fi\ifluatex 1\fi=0 % if pdftex
  \usepackage[utf8]{inputenc}
\else % if luatex or xelatex
  \usepackage{fontspec}
  \ifxetex
    \usepackage{xltxtra,xunicode}
  \fi
  \defaultfontfeatures{Mapping=tex-text,Scale=MatchLowercase}
  
\fi
\IfFileExists{microtype.sty}{\usepackage{microtype}}{}
\usepackage[]{natbib}
\ifxetex
  \usepackage[setpagesize=false, % page size defined by xetex
              unicode=false, % unicode breaks when used with xetex
              xetex]{hyperref}
\else
  \usepackage[unicode=true]{hyperref}
\fi
\hypersetup{breaklinks=true,
            pdfauthor={},
            pdftitle={Area- and unit-level small area estimation methods for multivariate outcome data},
            colorlinks=false,
            urlcolor=blue,
            linkcolor=magenta,
            pdfborder={0 0 0}}

\providecommand{\tightlist}{%
  \setlength{\itemsep}{0pt}\setlength{\parskip}{0pt}}

\usepackage{subcaption}
\usepackage{bm}

\newcommand{\nc}{\newcommand}

\nc{\E}{\mathbb{E}}
\nc{\Ehat}{\widehat{\mathbb{E}}}
\nc{\Var}{\mbox{Var}}
\nc{\Varhat}{\widehat{\mbox{Var}}}
\nc{\Cov}{\mbox{Cov}}
\nc{\Covhat}{\widehat{\mbox{Cov}}}
\nc{\R}{\mathbb{R}}
\nc{\pois}{\mbox{Poisson}}
\nc{\binomial}{\mbox{Binomial}}
\nc{\normal}{\mbox{Normal}}
\nc{\multinomial}{\mbox{Multinomial}}

\nc{\yij}{y_{ij}}
\nc{\yik}{y_{ik}}
\nc{\yijk}{y_{ijk}}
\nc{\yi}{\mathbf{y}_i}

\nc{\Y}{\mathbf{Y}}
\nc{\Yh}{\mathbf{Y}_h}

\nc{\bij}{b_{ij}}
\nc{\bi}{\mathbf{b}_i}

\nc{\betaij}{\beta_{ij}}
\nc{\betaj}{\bm{\beta}_j}

\nc{\alphaij}{\alpha_{ij}}
\nc{\alphaj}{\bm{\alpha}_j}

\nc{\xij}{\mathbf{x}_{ij}}

\nc{\mXi}{\mathbf{X}_i}
\nc{\mXk}{\mathbf{X}_k}
\nc{\mXh}{\mathbf{X}_h}
\nc{\mXkh}{\mathbf{X}_{kh}}

\nc{\muij}{\mu_{ij}}
\nc{\mui}{\bm{\mu}_i}
\nc{\nuij}{\nu_{ij}}
\nc{\nui}{\bm{\nu}_i}

\nc{\lambdaij}{\lambda_{ij}}
\nc{\lambdaj}{\lambda_{j}}
\nc{\lambdak}{\lambda_{k}}
\nc{\lambdaik}{\lambda_{ik}}
\nc{\lambdai}{\bm{\lambda}_i}

\nc{\lambdatildeij}{\tilde{\lambda}_{ij}}
\nc{\lambdatildeik}{\tilde{\lambda}_{ik}}
\nc{\lambdatildei}{\bm{\tilde{\lambda}}_i}
\nc{\Lambdai}{\bm{\Lambda}_i}
\nc{\Lambdatildei}{\bm{\tilde{\Lambda}}_i}

\nc{\D}{\mathbf{D}}

\nc{\dij}{d_{ij}}
\nc{\dik}{d_{ik}}
\nc{\dic}{d_{ic}}
\nc{\djk}{d_{jk}}
\nc{\djj}{d_{jj}}
\nc{\dkk}{d_{kk}}
\nc{\dikc}{d_{ikc}}
\nc{\dihkc}{d_{ihkc}}

\nc{\ti}{t_i}
\nc{\tik}{t_{ik}}

\nc{\deltai}{\delta_i}
\nc{\deltaij}{\delta_{ij}}
\nc{\deltaic}{\delta_{ic}}
\nc{\deltaik}{\delta_{ik}}
\nc{\deltaj}{\delta_j}
\nc{\deltaiA}{\delta_{iA}}
\nc{\deltaiB}{\delta_{iB}}

\nc{\LA}{\mathcal{L}^A}
\nc{\LB}{\mathcal{L}^B}
\nc{\LC}{\mathcal{L}^C}
\nc{\LS}{\mathcal{L}^S}
\nc{\Li}{\mathcal{L}_i}
\nc{\Lh}{\mathcal{L}_h}

\nc{\Lik}{\mathcal{L}_{ik}}
\nc{\LiA}{\mathcal{L}_i^A}
\nc{\LiB}{\mathcal{L}_i^B}
\nc{\LiC}{\mathcal{L}_i^C}
\nc{\LiS}{\mathcal{L}_i^S}

\nc{\tLA}{\tilde{\mathcal{L}}^A}
\nc{\tLB}{\tilde{\mathcal{L}}^B}
\nc{\tLC}{\tilde{\mathcal{L}}^C}
\nc{\tLS}{\tilde{\mathcal{L}}^S}
\nc{\tLi}{\tilde{\mathcal{L}}_i}

\nc{\tLiA}{\tilde{\mathcal{L}}_i^A}
\nc{\tLiB}{\tilde{\mathcal{L}}_i^B}
\nc{\tLiC}{\tilde{\mathcal{L}}_i^C}
\nc{\tLiS}{\tilde{\mathcal{L}}_i^S}

\nc{\tLiAone}{\tilde{\mathcal{L}}_i^{A,1}}
\nc{\tLiBone}{\tilde{\mathcal{L}}_i^{B,1}}
\nc{\tLiAtwo}{\tilde{\mathcal{L}}_i^{A,2}}
\nc{\tLiBtwo}{\tilde{\mathcal{L}}_i^{B,2}}

\nc{\tLij}{\tilde{\mathcal{L}}_i^j}

\nc{\barlambda}{\overline{\lambda}}

\nc{\bz}{\bm{z}}

\nc{\nqx}[2]{\tensor*[_{#1}]{q}{_{#2}}}
\nc{\nqxc}[3]{\tensor*[_{#1}]{q}{^{#3}_{#2}}}

\newcommand{\eq}[1]{\begin{align*}#1\end{align*}}
\newcommand{\eqnum}[1]{\begin{align}#1\end{align}}

\newcommand{\blandscape}{\begin{landscape}}
\newcommand{\elandscape}{\end{landscape}}

\usepackage{xcolor}

\definecolor{gg1}{HTML}{F8766D}
\definecolor{gg2}{HTML}{C49A00}
\definecolor{gg3}{HTML}{53B400}
\definecolor{gg4}{HTML}{00C094}
\definecolor{gg5}{HTML}{00B6EB}
\definecolor{gg6}{HTML}{A58AFF}
\definecolor{gg7}{HTML}{FB61D7}
\definecolor{gg8}{HTML}{C77CFF}

\usepackage{booktabs}
\usepackage{longtable}
\usepackage{array}
\usepackage{multirow}
\usepackage{wrapfig}
\usepackage{float}
\usepackage{colortbl}
\usepackage{pdflscape}
\usepackage{tabu}
\usepackage{threeparttable}
\usepackage{threeparttablex}
\usepackage[normalem]{ulem}
\usepackage{makecell}
\usepackage{xcolor}

\begin{document}

\begin{frontmatter}

  \title{Small Area Estimation Methods for
Multivariate Health and Demographic Outcomes using Complex Survey Data}
    \author[UW HMS]{Austin E Schumacher%
  \corref{cor1}%
  }
   \ead{aeschuma@uw.edu} 
    \author[UW Biostat Stat]{Jon Wakefield%
  }
   \ead{jonno@uw.edu} 
      \affiliation[UW HMS]{
    organization={Department of Health Metrics Sciences, University of
Washington},addressline={3980 15th Ave
NE},city={Seattle},postcode={98195},state={WA},country={United States},}
    \affiliation[UW Biostat Stat]{
    organization={Department of Biostatistics, Department of Statistics,
University of Washington},addressline={3980 15th Ave
NE},city={Seattle},postcode={98195},state={WA},country={United States},}
    \cortext[cor1]{Corresponding author}
  
  \begin{abstract}
  Improving health in the most disadvantaged populations requires
  reliable estimates of health and demographic indicators to
  inform policy and interventions. Low- and middle-income countries with
  the largest burden of disease and disability tend to have the least
  comprehensive data, relying primarily on household surveys. 
  Subnational estimates are increasingly used to inform targeted
  interventions and health policies. Producing
  reliable estimates from these data at fine geographical scales 
  requires statistical modeling, and small area estimation models
  are commonly used in this context. Although most
  current methods model univariate outcomes, improved
  estimates may be attained by borrowing strength across related
  outcomes via multivariate modeling. In this paper, we
  develop classes of area- and unit-level multivariate shared component
  models using complex survey data. This framework jointly models
  multiple outcomes to improve accuracy of estimates compared to
  separately fitting univariate models. We conduct simulation studies to
  validate the methodology and use the proposed approach on survey data
  from Kenya in 2014; first, to jointly model height-for-age  and
  weight-for-age in children, and second, to model three categories of
  contraceptive use in women. These models produce improved estimates
  compared to univariate and naive multivariate modeling approaches.
  \end{abstract}
  
 \end{frontmatter}

\hypertarget{intro}{%
\section{Introduction}\label{intro}}

Reliable estimates of health and demographic indicators in low- and
middle-income countries (LMICs) are of paramount importance to describe the health and
developmental landscape of the places in most need, serving to expose
inequalities that can be addressed by policy and interventions. In many
LMICs, household surveys such as the Demographic and
Health Surveys (DHS) program \citep{corsi2012demographic} are
the primary source of nationally-representative data \citep{boerma2007health}. 
These data are often analyzed using small area estimation (SAE), an umbrella 
term for statistical methods that are used to produce estimates for domains 
with sparse outcome data. \citet{rao2015small} is the standard
reference on SAE, and \citet{pfeffermann2013new} gives an in-depth
methodological review. \citet{wakefield2020small} provides an overview
specific to health outcomes in LMICs.

Using data from complex surveys requires accounting for the
survey sampling design, which is an important and sometimes overlooked
component of SAE (in the spatial statistics literature, see \cite{wakefield25two}). 
Acknowledging survey design in SAE has led to
\emph{design-based} and \emph{model-based} approaches 
\citep{skinner2017introduction}. This paper focuses on model-based
approaches. Methods are split into two types:~\emph{area-level}, for 
which covariate information and model parameters
are specified at the desired level of aggregation,
and \emph{unit-level}, for which data are modeled at the level of the
individual or cluster. For area-level models, the seminal Fay-Herriot
model \citep{fay1979estimates} uses a weighted estimate as the outcome 
and random effects for each area. Many extensions of the Fay-Herriot model 
have been proposed, see \citet{rao2015small} and 
\citet{you2011hierarchical, marhuenda2013small, mercer2015space, watjou2017spatial}.
One of the first unit-level approaches was a nested error
regression model proposed by \citet{battese1988error}, which has had
various extensions, such as 
\cite{macgibbon1989small} and \cite{ghosh1998generalized}. See 
\citet{rao2015small} and \citet{parker2019unit} for  comprehensive reviews. 

Deciding between design- or model-based approaches, and between 
area- and unit-level models, can be tricky. 
\citet{paige2022design} conclude that direct estimators are
the gold standard when sufficient data allow for an acceptable level of
uncertainty, while model-based area-level estimators can aid in
reducing variance, albeit with the introduction of bias due
to smoothing; but when direct estimates are unreliable,
unit-level models are required, which entails using a model that is
consistent with the survey design and requires careful effort to
specify a model. \citet{wakefield25two} provides a recent review of
SAE that compares direct estimators, area-level, and unit-level models,
similarly concluding that sufficient power for direct estimation
requires sufficiently large sample sizes, while model-based 
estimation---unit-level models in particular---require many layers 
of assumptions and a difficult aggregation step if unit-level covariates 
and/or continuous spatial models are used.

Most work on SAE involves single outcomes, which has seen the bulk of
recent developments in area-level models
\citep{ybarra2008small, gonzalez2010small, jiang2011best, datta2011estimation, datta2011bayesian, esteban2011area, herrador2011fay, slud2011small, kubokawa2012measuring, ghosh2013two, bell2013benchmarking, pfeffermann2014single}
and unit-level models
\citep{hobza2018small, diggle2019model}. However, the
scientific question of interest may involve multiple related outcomes,
or we may wish to use information from multiple outcomes to improve the
accuracy of predictions. In these settings, jointly modeling multiple
outcomes allows for borrowing of strength. Some examples include modeling 
height and weight for age, cause-specific mortality, neonatal mortality 
and female educational attainment, and multiple poverty measures. 
Many multivariate area-level SAE methods exist
\citep{fay1987application, fuller1987multivariate, datta1991hierarchical, datta1996estimation, datta1998multivariate, datta1999empirical, huang2006using, gonzalez2008analytic, fay2013small}.
\citet{benavent2016multivariate} surveys this literature while also
proposing a multivariate Fay-Herriot model. More recent developments in 
multivariate Fay-Herriot models can be found in \cite{arima2017multivariate, saegusa2020parametric, franco2021using, erciulescu2022model}. For multivariate unit-level SAE models, much
work has focused on multinomial data
\citep{ghosh1998generalized, molina2007small, saei2012labour, wang2018small, zhang2004small, berg2014small,bradley2018computationally,bradley2019spatio} and multiple
binary outcomes \citep{lawson2020multi}.
Several models have also been developed to jointly estimate multiple continuous outcomes 
\citep{lohr2003small, ngaruye2017small, ito2021empirical, li2010using, sun2021bivariate, esteban2022small}.

SAE methods aim to produce reliable estimates for small areas by
borrowing strength via auxiliary information---one common method uses
spatial models to exploit similarities across geographic areas. Spatial
models are categorized as either discrete or continuous. Much research has 
been done on multivariate spatial modeling in contexts other than SAE
\citep{gelfand2003proper, carlin2003hierarchical, jin2005generalized, carlin2007bayesian, jin2007order, neelon2014multivariate}.
One method in particular for discrete spatial analysis is the shared
component model, popularized by \citet{knorr2001shared}, whereby some 
parameters of the model are shared among multiple outcomes. 
\citet{macnab2010bayesian} proposes a general formulation of shared component
models that allows errors in covariates. Importantly, a
bivariate shared component model can be interpreted as an
ecological regression model, which allows the
coefficient for the latent shared component to be interpreted as the
regression effect of unobserved covariates \citep{knorr2001shared}. Since 
we often believe survey data have unobserved covariates, 
shared component models are an attractive option. 
Recent research has explored multivariate spatial area-level SAE without 
investigating shared component models \citep{porter2015small, guha2021measuring}.

At the unit-level, multivariate spatial modeling has been extensively developed. 
Coregionalization, in which latent parameters are shared among outcomes, is a 
popular approach and typically uses continuous spatial models
\citep{schmidt2003bayesian, palmi2019bayesian}. \citet{macnab2016linear}
contains a thorough review. Coregionalization is essentially the
continuous spatial analog of discrete shared component models.
Continuous spatial models naturally arise for unit-level analyses
because they allow specification of spatial effects at the same
resolution as the observed data. However, if area-level estimates are
desired, aggregation is challenging because it requires auxiliary population
information. This can be problematic when the models use
non-linear link functions. In addition, the fine-grained population
densities required are typically modeled quantities with uncertainty,
and incorporating this uncertainty into the aggregate estimates is
difficult. 

%This motivates the use of
%discrete-space shared component models in unit-level analyses. Discrete
%spatial models for unit-level data have been used in the univariate case
%\citep{wakefield2020small}, but a gap in the SAE literature persists for
%spatial modeling at the unit-level using complex survey data for
%multivariate outcomes. The frontier of this research has considered
%multinomial outcomes \citep{parker2022computationally}, and multiple
%binary outcomes \citep{lawson2020multi}. However, the %general multivariate
%case has not been thoroughly explored, and shared
%component models have not been applied.

This paper develops novel multivariate SAE approaches for complex survey data.  
We propose classes of area- and unit-level shared component models with 
discrete spatial effects that acknowledge the survey design and conduct 
simulation studies to validate the methodology. We present two motivating examples 
using data from the 2014 Kenya Demographic and Health Survey (KDHS) \citep{kenya2015kenya}: 
jointly modeling height for age and weight for age in children under age 5, presented 
in the main text, and modeling three categories of contraceptive use in women aged 15--49, 
presented in Appendix A. Using the proposed modeling approaches, we produce
estimates for these examples that outperform comparison results from univariate models 
and multivariate non-shared models.

\hypertarget{data}{%
\section{Data Description}\label{data}}

The DHS Program, in collaboration with many LMICs, conducts
nationally-representative household surveys that provide data for a wide
range of monitoring and impact evaluation indicators in the areas of
population, health, and nutrition. The DHS Program uses a set of
consistent sampling approaches from country to country, with methods
described in the 2012 DHS Sampling and Household Listing Manual
\citep[Sec. 5.2, p.~80--85]{demographic2012health}. The standard design
is a stratified two-stage cluster sampling scheme with stratification by
region crossed with urban/rural. This article uses data 
from the 2014 KDHS. Appendix B provides 
a detailed description of the data. 

As a motivating example for jointly modeling continuous outcomes, we
will use two key child health indicators: height for age z-scores (HAZ)
and weight for age z-scores (WAZ). These are measures of growth in 
children under age five, with high values being preferable. 
We expect these variables to be correlated---at the individual 
level, cluster level, and area level---due to shared unobserved risk factors at
each of these levels. HAZ and WAZ have previously been jointly modeled from
survey data in Papua New Guinea using multivariate CAR models with spatial
random effects \citep{gelfand2003proper}; however, this analysis did not 
account for the survey sampling design. 

We present the region-level univariate naive unweighted estimates (not
accounting for the survey design) and weighted estimates
(accounting for the survey design) in Figure 1. Compared to HAZ, 
WAZ displays a more consistent north (lower) to south (higher) spatial gradient. 
The urban areas in south-central Kenya tend to have the highest HAZ and WAZ. However, 
the standard errors of the univariate weighted (direct) estimates
are fairly wide, especially in the more rural counties, and the
levels are different between HAZ and WAZ. Ignoring the survey design
gives national mean estimates (standard error) of -1.00 (0.0099) for HAZ
and -0.89 (0.0089) for WAZ, while the weighted estimates are -0.96
(0.016) and -0.78 (0.016), respectively. The increased standard error
is due to clustering. Figure 2 presents a scatterplot of the
region-level design-weighted direct estimates of HAZ versus WAZ with
uncertainty, showing a positive association between the two measures 
along with similar uncertainty intervals.

\begin{figure}

{\centering \subfloat[Mean estimates\label{fig:univariate-1}]{\includegraphics[width=0.48\linewidth]{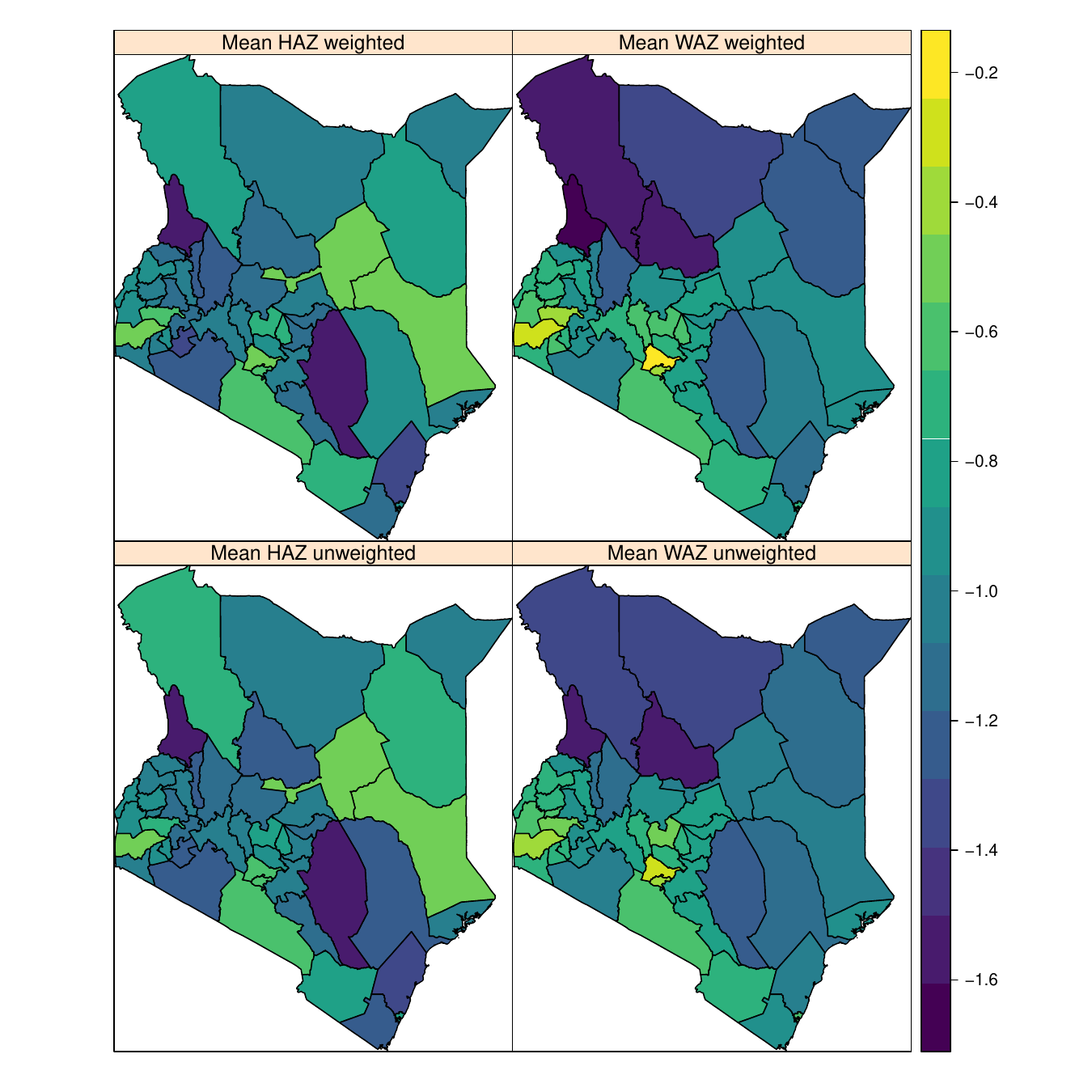} }\subfloat[Standard errors\label{fig:univariate-2}]{\includegraphics[width=0.48\linewidth]{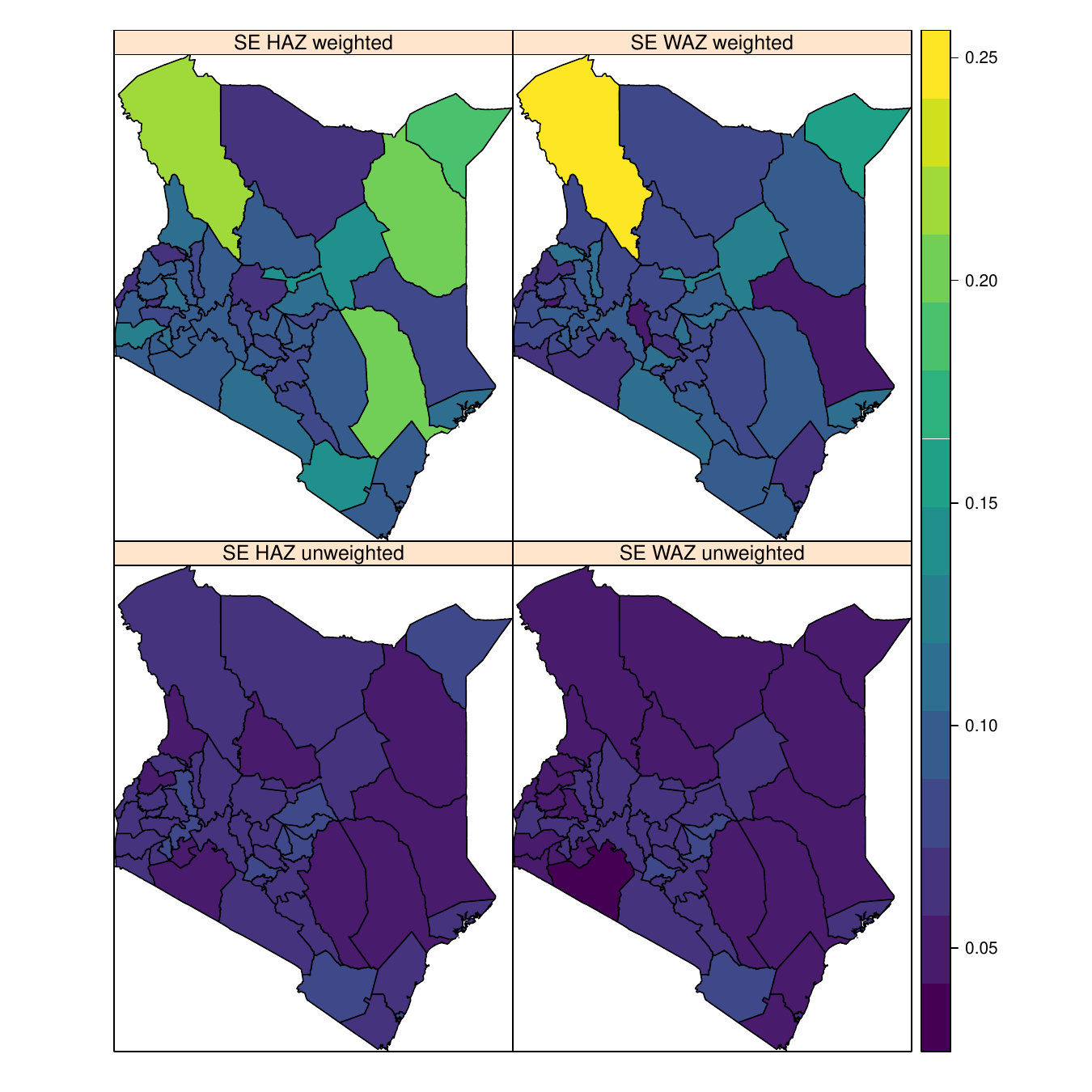} }

}

\caption{Univariate direct estimates of means (left two columns) and standard errors (right two columns) for HAZ and WAZ from the 2014 KDHS. We present both naive unweighted estimates that do not account for the complex survey design, and weighted estimates that do.}\label{fig:univariate}
\end{figure}

\begin{figure}

{\centering \includegraphics[width=0.45\linewidth]{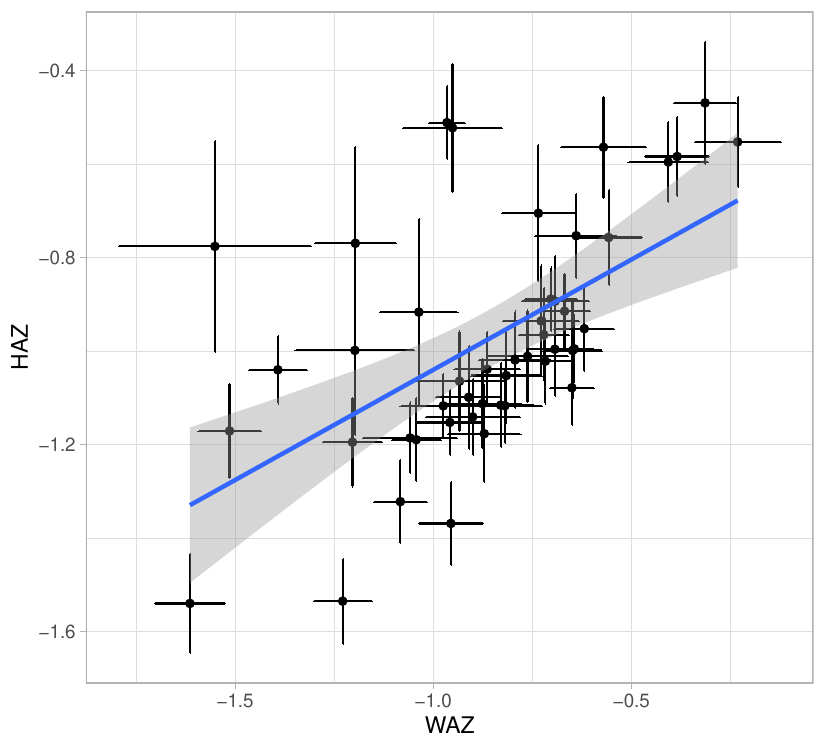} 

}

\caption{Weighted univariate direct estimates of mean HAZ verses WAZ at the region level. Error bars represent plus/minus one standard error. Simple linear regression line with 95\% CI is included.}\label{fig:haz-waz-scatter}
\end{figure}

As a motivating example for multinomial data, we will model current
contraceptive use in women aged 15-49 from the 2014 KDHS. Modeling the data as multinomial, rather than modeling one category at a time,
is important due to the implicit dependencies among multinomial
probabilities---they must sum to 1. For contraceptive use in particular,
different methods may be used by similar subgroups that have shared,
unobserved characteristics in different areas, e.g., due to cultural
practices or education. This is further motivation for joint modeling.
We grouped the 14 possible categories of contraceptive use in this survey 
into three broad categories to have sufficient observations in each:
\emph{none} (no contraceptive use), \emph{modern} (pill, IUD,
injections, condom, female sterilization, male sterilization,
implants/norplant, lactational amenorrhea, female condom, and other
modern methods), and \emph{other} (periodic abstinence, withdrawal, and
other).

\hypertarget{methods}{%
\section{Methods}\label{methods}}

%The outcome, time period, and geographic level of interest for
%demographic and health indicators in the DHS and similar surveys tend 
%to not have sufficient data for direct estimates to be adequately precise, yet
%they do have enough data for reliable model-based variance estimates. Thus,
%model-based methods are of particular importance. 

In this section, we propose shared component modeling frameworks for
multivariate outcomes. First, we describe spatial smoothing methods
and how they are used in shared component modeling. We then discuss
area-level modeling, for which we develop a Fay-Herriot type two-stage
approach, before moving to unit-level modeling, for which we develop an
approach using discrete spatial effects and shared latent parameters
among outcomes. We describe the models for continuous outcomes, and in 
Appendix A, we adapt them for multinomial outcomes. We take a Bayesian
approach to modeling in this paper, primarily due to the 
availability of fast, accurate, and user-friendly software implementations 
that we will describe in Section 3.5. 
% We first briefly describe the shared component spatial model, since this 
% is common to both the area-level and unit-level models we describe.

\hypertarget{spatial-modeling}{%
\subsection{Spatial shared component models}\label{spatial-modeling}}

Spatial smoothing methods are commonly used to model demographic and
health variables from survey data in LMICs \citep{wakefield25two}. In
particular, we focus on discrete spatial models.
\citet{banerjee2014hierarchical} provides a review of choices for
modeling the spatial random effects. In this paper, we model area-level
spatial effects with a Besag, York and Mollié (BYM) model which includes
both an intrinsic conditional autoregressive (ICAR) component for
spatial auto-correlation and an independent and identically distributed
(IID) normal component for non-spatial heterogeneity
\citep{besag1991bayesian}. %In the univariate case, the BYM model can be
%written as \(v_{r} + u_{r}\), with
%\(v_{r} \mid  \sigma^2_{v} \sim_{iid} N(0, \sigma^2_{v})\) the
%nonspatial IID normal random effects on region, and
%\(\bm{u} \mid  \sigma^2_{u} \sim ICAR(\sigma^2_{u})\) the ICAR spatial
%random effects are given conditional representation,
%$$u_{r} \mid  u_{r'}, r' \neq r \sim N\left(\frac{1}{n_r}\sum_{r \in S_r}  u_{r}, \frac{\sigma^2_{u}}{n_r}\right),$$
%where $S_r$ is the set of neighbors of region $r$, which in this paper we take to be the set of regions that share a boundary, and $n_r$ is the number of such neighbors.
We use the parameterization of the BYM model from \citet{riebler2016intuitive} 
that reformulates the IID and ICAR random effects in area $r$ as
\eq{
    v_{r} + u_{r} = \sigma(\sqrt{1 - \rho}v_{r}^* + \sqrt{\rho}u_{r}^*).
}

In this formulation, $\sigma$
%= \sqrt{\sigma^2_{v} + \sigma^2_{u}}\) 
is the total standard
deviation of the random effects. The unstructured component is
\(v_{r}^* \sim_{iid} N(0,1)\). The spatial ICAR component is \(u_r^*\), 
and these are scaled so \(\mbox{Var}(u_{r}^*) \approx 1\). To do this, we
follow \citet{riebler2016intuitive} and scale the model so that the geometric 
mean of the variances is 1, using the adjacency matrix to calculate the 
inverse precision of the ICAR model. The parameter \(\rho\) is interpreted
as the proportion of the total variation that is due to the spatial random
effect. This version of the BYM model is an improvement over the original
because the parameters are more interpretable, allowing more intuitive specification 
of prior distributions.

Extending discrete spatial models to the multivariate setting is crucial
to leverage shared spatial information between outcomes. The simplest
case would be to specify separate BYM effects for each outcome.
However, it is plausible that many demographic and health outcomes share common but
unobserved covariates. This naturally leads to models in the literature
called shared component models, for which latent parameters are shared
among outcomes \citep{knorr2001shared}. These models can be rewritten as 
ecological regression models with errors in covariates, which is attractive
as this reflects the nature of demographic and health outcomes having 
shared, unobserved covariates (described in Appendix B). Using two outcomes $c = 1, 2$ 
for simplicity, we formulate the shared component model as
\eq{
    \mu_{r1} &= \beta_1 + s_{r1} + \lambda s_{r2} \\
    \mu_{r2} &= \beta_2 + s_{r2},
}
\noindent with BYM effects $s_{rc} = v_{rc} + u_{rc} = \sigma_c(\sqrt{1 - \rho_c}v_{rc}^* + \sqrt{\rho_c}u_{rc}^*)$ 
for each outcome \(c\).

Theoretically, a parameterization with a shared $s_{r2}$ component 
is equivalent to a parameterization with a shared $s_{r1}$ component, 
as we demonstrate in Appendix B. In practice, some differences may arise 
from numerical instability, particularly in Bayesian models and uncertainty estimation. 
We explore this further in the contraceptive use example (Appendix A). 

Our chosen specification is different from the one proposed in \citet{knorr2001shared}, 
which includes an additional outcome-specific random effect in $\mu_{r2}$. 
Although this additional random effect is appealing because it yields a symmetric 
formulation, issues with identifiability arise without strong priors on the 
precision parameters or additional constraints on the random effects.

\subsection{Area-level modeling framework}\label{modeling-frameworks}

We develop a multivariate shared component Fay-Herriot model, which is a
two stage modeling approach. In the first-stage, we develop a working
likelihood for the multivariate outcomes to obtain a region-level summary. In the second
stage, we fit a mixed effects smoothing model for the region-level
estimates produced at the first stage.

\subsubsection{Area-level modeling notation}\label{notation}

For area-level models, assume the \(R\) regions are the units of
analysis. Let \(Y_{ric}\) be an observation of outcome
\(c = 1, \dots, C\) for individual \(i = 1, \dots, N_r\) in region $r$, with \(N_r\) the number of individuals in region
\(r\) (population size), $r = 1, \dots, R$. Often, the population mean in area \(r\) is of
particular interest, \(\bar{Y}_{rc} = \sum_{i=1}^{N_r} Y_{ric}/ N_r\).
In the univariate case ($C = 1$), outcomes \(Y_{ri1}\) are typically
binary or continuous in the DHS surveys we focus on. In the multivariate case, \(Y_{ric}\) typically
refer to multiple continuous variables or observations of a categorical
variable with \(C\) categories.

A survey is conducted to obtain a sample of \(n_r\) individuals in each
region, a subset of the population. Let \(I_{ri}\) be an indicator for
membership into the sample. Define \(\pi_{ri}\) to be the first-order
inclusion probability, which is the probability that individual \(i\) in
area \(r\) is selected, i.e., \(\Pr( I_{ri} = 1) = \pi_{ri}\), and
\(y_{ric}\), $c=1,\dots,C$, are the observed outcomes associated with this sampled individual. Define the survey design
weights as \(w_{ri} = 1 / \pi_{ri}\). 
% Further define \(\pi_{rij}\) to be the
% second-order inclusion probability, which is the probability that both
% individuals \(i\) and \(j\) are selected in area \(r\). 
% Throughout this
% paper, we will use bold font and drop subscripts to refer to vectors.
The vector of all observed outcomes for individual \(i\) in region
\(r\) is denoted \(\bm{y}_{ri} = [y_{ri1}, \dots, y_{riC}]\).

\subsubsection{Area-level model
description}\label{area-level-model-description}

In the first stage of our modeling approach, we
take the individual-level outcomes and calculate survey weighted
averages in each region that account for 
the complex survey design, along with a design-based estimated
covariance matrix for these means. This procedure results in a vector 
of the survey-weighted mean outcomes for each region, as well as a
design-based estimate for the covariance of this random vector.

% First, for each area we will calculate a vector of means of each outcome
% and its associated survey design-based covariance matrix using weighted
% estimates and the appropriate variance.
To produce mean estimates in each region, we calculate the Hájek estimator as
\eq{
    \hat{\bar{y}}_{rc} &= \frac{\sum_{i=1}^{n_r} w^{*}_{ri} y_{ric}}{\sum_{i=1}^{n_r} w^{*}_{ri}}.
}
\noindent with weights $w^{*}_{ri}$ normalized by their sum. We then calculate the estimated 
design-based covariance matrix of
\(\hat{\bar{\bm{y}}}_r = (\hat{\bar{y}}_{r1}, \dots, \hat{\bar{y}}_{rC})\)
for each region
as
\eq{
    \hat{\bm{V}}^{des}_{r} &= \begin{bmatrix} \hat{V}^{des}_{r11} & \hat{V}^{des}_{r12} & \dots & \hat{V}^{des}_{r1C} \\ \hat{V}^{des}_{r21} & \hat{V}^{des}_{r22} & \dots & \hat{V}^{des}_{r2C} \\ \vdots & \vdots & \ddots & \vdots \\ \hat{V}^{des}_{rC1} & \hat{V}^{des}_{rC2} & \dots & \hat{V}^{des}_{rCC} \\ \end{bmatrix}.
}
\noindent This estimate must account for the stratified cluster
survey design. The form of \(\hat{\bm{V}}^{des}_{r}\) and its
derivation are provided in Appendix B. These direct
estimates and design-based covariance matrices are estimated separately
for each region.

For the second stage, we fit a Bayesian smoothing model on these
region-specific estimates that uses the mean vectors from the first stage,
\(\hat{\bar{\bm{y}}}_r\), as the outcome, with associated variance-covariance matrix,
\(\hat{\bm{V}}^{des}_{r}\). The smoothing model can be flexibly
specified depending on the available data and the scientific question of
interest; we envision that it will typically be a mixed effects model
including area-level covariates along with spatial random effects.

As with all Fay-Herriot models, the sampling distribution for 
$\hat{\bar{\bm{y}}}_r$ is taken as the asymptotic sampling distribution 
of the estimator:
% Due to the amount of data, we will assume that the mean outcomes in a
% region have a multivariate normal distribution with variance equal to
% \(\hat{\bm{V}}^{des}_{r}\). We propose models with the working
% likelihood
\eqnum{
    \hat{\bar{\bm{y}}}_{r} \mid  \bm{\mu}_{r},  \hat{\bm{V}}^{des}_{r} &\sim N_2(\bm{\mu}_{r},  \hat{\bm{V}}^{des}_{r}) \label{p2-eq-likelihood}
}

\noindent with \(\hat{\bar{\bm{y}}}_{r}\) playing the role of the
``observed'' data and  with \(\hat{\bm{V}}^{des}_{r}\) assumed fixed and
known. Then, we specify a mean smoothing model on the latent parameters
\(\bm\mu_r\). The smoothing models that we investigate in this paper
will include BYM random effects to leverage spatial information, and
will also be formulated as shared component models (Section 3.1) to borrow
strength among outcomes.

\subsection{Unit-level modeling framework}\label{unit-level-modeling-multivariate-discrete-spatial-model-with-shared-latent-parameters}

We develop a multivariate discrete spatial model with shared latent
parameters. This model uses fixed and random effects specified with
hierarchical structure.

\subsubsection{Unit-level modeling notation}\label{notation-1}

For unit-level models, assume the units of analysis are clusters within
a multistage cluster design. Let \(i = 1, \dots, N_k\) denote
individuals which are from \(k = 1, \dots, K_r\) clusters sampled from
\(r = 1, \dots, R\) regions, and let \(c = 1, \dots, C\) denote outcomes.
%Let \(\bm{s}_{rk}\) represent the geographical location of cluster \(k\) in region \(r\). 
Let \(Y_{rkic}\) be the random variable for outcome
\(c\) of individual \(i\) in cluster \(k\) of area \(r\), with observed
value \(y_{rkic}\). 
% The vector of observed outcomes for individual \(i\)
% in cluster \(k\) of region \(r\) is denoted
% \(\bm{y}_{rki} = [y_{rki1}, \dots, y_{rkiC}]\).

\subsubsection{Unit-level model
description}\label{unit-level-model-description}

We use a Gaussian likelihood for continuous outcomes. For outcome
\(c\) of individual \(i\) in cluster \(k\) of region \(r\), we
specify the \emph{individual-level} likelihood,
\eq{
    Y_{rkic} \mid  \mu_{rkic}, \omega^2_c &\sim N(\mu_{rkic}, \omega^2_c).
}
\noindent which has outcome-specific, individual-level variances
\(\omega^2_c\). In practice, we may allow these variances to differ by region $r$. Of note, we do not explicitly use a multivariate
likelihood across the outcomes, but instead we let the shared components
induce dependency among outcomes, which is similar to continuous
coregionalization models \citep{schmidt2003bayesian}.

% AUSTIN: Should we have a covariance here, beyond that induced by the spatial model?
% JON: I think we could, but for simplicity I like having all of the correlation specified by the shared components... That's why I state "Of note, we do not explicitly use a multivariate likelihood across the outcomes, but instead we let the shared components induce dependency among outcomes..." Including additional covariance would mean having to estimate it, which would be really cumbersome.

The discrete spatial shared component model for \(\mu_{rkic}\) must be
developed with regard to the survey design used to sample the
observed data. Considering the stratified cluster design of the 2014 KDHS
(along with most other DHS surveys), for which the strata consist of the first
administrative regions crossed with the binary classification of
urban or rural, we must include effects in our model for both regions and 
urban/rural classification. Due to our desire
to include spatial random effects, we do not use fixed intercepts for
strata because that would cause identifiability problems---rather, we
use random effects on regions, along with fixed effects on
urban/rural indicators, for each outcome. This imposes a simplifying assumption of constant
urban/rural association across regions. Considering two outcomes for simplicity,
\(c = 1, 2\), the latent parameters are modeled as
\begin{align*}
    \mu_{rki1} &= \beta_1 + z_{rk} \gamma_1  + s_{r1} + \lambda s_{r2} + \epsilon_{rk1} \\
    \mu_{rki2} &= \beta_2 + z_{rk} \gamma_2  + s_{r2}  + \epsilon_{rk2}.
\end{align*}

We have outcome-specific fixed intercepts, \(\beta_c\), while
\(z_{rk}\) is the binary indicator of whether cluster \(k\) in region \(r\)
is rural (rather than urban), and \(\gamma_c\) is the
outcome-specific effect associated with rural areas. For the discrete spatial component, 
$s_{rc}$ are BYM effects. For the shared component, we include the BYM random effects
from outcome \(c = 2\) in the parameterization for the mean of outcome
\(c = 1\). These are scaled by a coefficient \(\lambda\) to be estimated
from the model. This is a proxy for unobserved shared covariates between 
outcomes.

The cluster-level errors are
\(\epsilon_{rkc} \mid  \sigma_{\epsilon,c}^2 \sim_{iid} N(0,\sigma_{\epsilon,c}^2)\).
We consider these to represent measurement error in this
context, and so we do not include these terms when making model-based
predictions. We will assume no covariates for simplicity, though this
model can include them with (outcome-specific) fixed
effects. Surveys usually collect covariates that we may wish to use, 
but the aggregation step from clusters to areas needs careful consideration 
\citep{wakefield25two}. While linear models offer simplifications, 
area-level means are still required. This is often difficult in LMICs due to 
a lack of reliable or recent censuses.

In order to calculate area-level means, we aggregate the model over 
urban/rural classification to give
\eq{
    \mu_{rc} &= (1 - q_r) \times \big( \beta_c + g_c \big) + q_r \times \big( \beta_c + \gamma_c + g_c \big).
}
\noindent Here, \(g_c\) are the (shared) random effects
corresponding to outcome \(c\), and \(q_r\) are the proportion of the
relevant population in region \(r\) that is rural. For example, when
modeling HAZ and WAZ in children under 5, \(q_r\) is the
proportion of the under-5 population in region \(r\) that live in rural
areas.

\hypertarget{model-selection}{%
\subsection{Model selection}\label{model-selection}}

Model selection among candidate models is an important task, but it 
is not straightforward in the context of model-based SAE. We use 
distinct model selection approaches for area- and unit-level models.

\hypertarget{area-level-model-selection}{%
\subsubsection{Area-level model
selection}\label{area-level-model-selection}}

For Fay-Herriot models, the first stage can be considered as a
``data-processing'' step in some sense.
%, as the performance of the first
%stage direct estimates and covariance matrix have been validated in
%previous research \citep{mercer2015space}. Thus 
We will take the first stage direct estimates as if they were observed data and
perform model selection for a suite of candidate models at the second
stage. Models will be compared via leave-one-out cross validation using
a multivariate scoring strategy that accounts for the correlation
between the outcomes, e.g. HAZ and WAZ in our primary example. 
For region \(r\), we define
\begin{equation}\label{eq:pred}
\ell_r = p(\hat{\bar{\bm{y}}}_r \mid  \bm{y}_{-r}) = \int_{\mu_r} p( \hat{\bar{\bm{y}}}_r \mid \bm{\mu}_r ) \times p(\bm{\mu}_r \mid
\bm{y}_{-r}) ~d\bm{\mu}_r
\end{equation}
as the posterior predictive distribution from a model fit with region \(r\) removed evaluated at the
true value of the held out data---in this case, the true value of the
held out data are the direct estimates of the held out region, $\hat{\bar{\bm{y}}}_r$. We
perform the following steps for each region:
\begin{enumerate}
\def\labelenumi{\arabic{enumi}.}
\tightlist
\item
  Hold out the direct estimates for region \(r\), $\hat{\bar{\bm{y}}}_r$, $r=1,\dots,R$.
\item
  Fit all candidate models to the direct estimates with region \(r\)
  removed.
\item
  Draw samples from the posterior distribution,
  \(\bm{\mu}_{r}^{(s)} = p(\bm{\mu}_r \mid  \bm{y}_{-r})\), $s=1,\dots,S$, with $S=1,000$.
\item
  Calculate
  \(\hat\ell_r = \frac{1}{S} \sum_{s = 1}^{S} p(\hat{\bar{\bm{y}}}_r \mid  \bm{\mu}_{r}^{(s)}, \bm{\hat{V}}^{des}_{r})\),
  where
  \(p(\hat{\bar{\bm{y}}}_r \mid  \bm{\mu}_{r}^{(s)}, \bm{\hat{V}}^{des}_{r}) = N_2(\bm{\mu}_{r}^{(s)}, \bm{\hat{V}}^{des}_{r})\), as an approximation to (\ref{eq:pred}).
\item
  Calculate \(- \sum_{r = 1}^R \log(\hat \ell_r)\), which we will call the
  \emph{LogScore}. 
\end{enumerate}

We will compare LogScore statistics across models. Since the LogScore is based
on the negative log-likelihood, lower values indicate better performing
models.

% AUSTIN: Why did we take the -ve? Isn't it easier to interpret when these are large?
% JON: I did this in order to be consistent with the definition of scoring rules. From Wikipedia, "Scoring rules that are (strictly) proper are proven to have the lowest expected score if the predicted distribution equals the underlying distribution of the target variable. Although this might differ for individual observations, this should result in a minimization of the expected score if the "correct" distributions are predicted."

\hypertarget{unit-level-model-selection}{%
\subsubsection{Unit-level model
selection}\label{unit-level-model-selection}}

Area-level means remain the target of estimation for unit-level models.
Thus, we design a similar procedure that validates at the area-level 
utilizing the predictive distribution of the true area-level mean vectors,
\(\bm{\mu}_{r}\).

% If we hold out all data in region \(r\) when fitting
% the model, the direct estimate of the held out data (i.e.~the Hájek
% estimator
% \(\hat{\bar{\bm{y}}}_{r} \sim N_C(\bm\mu_r, \hat{\bm{V}}^{des}_r)\) with
% \(\hat{\bm{V}}^{des}_r\) the design-based covariance matrix) is
% independent of this predictive distribution from the model fit with
% region \(r\) removed because it is based on independent data. Thus,
% combining the independent sampling distribution of the direct estimator
% for the held out region with the posterior for the true value from the
% model fit with the held out data removed, we find a predictive
% distribution for the held out data that we can use to validate estimates
% from the model at the area-level.
For each region \(r\), the procedure follows steps 1--5 above, except step 3 is replaced by
\begin{enumerate}
\def\labelenumi{\arabic{enumi}.}
\tightlist
% \item
%   Hold out all observations for region \(r\).
% \item
%   Fit all candidate models to the data with region \(r\) removed.
\item[3.]
  Sample \(s = 1, \dots, S\) samples from the posterior distributions of
  the region-specific bivariate means,
  \(\bm\mu_{r}^{(s)} = [\mu_{r1}^{(s)}, \mu_{r2}^{(s)}]\), where
  \(\mu_{rc}^{(s)}\) are the weighted averages of the urban and rural
  estimates,
  i.e.,
  $$\mu_{rc}^{(s)} = (1 - q_r) \times \big( \beta_c^{(s)} + g_c^{(s)} \big)  + q_r \times \big( \beta_c^{(s)} + \gamma_c^{(s)} + g_c^{(s)} \big) 
  $$
  with the \(\beta_c^{(s)}\), \(\gamma_c^{(s)}\), and \(g_c^{(s)}\)
  parameters as defined previously with the \((s)\) superscript denoting
 posterior sample \(s\). 
 % Remember, these samples are
 %  from models fit to data with region \(r\) removed.
% \item
%   Compute the survey-weighted direct estimates of the means for the held
%   out data, \(\hat{\bar{\bm{y}}}_{rc}\), along with the appropriate
%   design-based covariance \(\hat{\bm{V}}^{des}_r\). See Appendix B for
%   details on this calculation.
% \item
%   For each sample \(s\), the posterior predictive distribution of the
%   held out data is \(N_C(\bm\mu_{r}^{(s)}, \hat{\bm{V}}^{des}_r)\), and
%   we evaluate this distribution at the value
%   \(\hat{\bar{\bm{y}}}_{rc}\). We call these likelihood values
%   \(\tilde\pi^{(s)}_r\).
% \item
%   Average these over all samples and calculate the negative
%   loglikelihood,
%   \(\ell_r = -\log\left(\frac{1}{S}\sum_{s = 1}^S \tilde\pi^{(s)}_r \right)\).
\end{enumerate}

% \noindent Once we have all of these region-level scoring estimates, we
% average over them and end up with

% \eq{
%     \mbox{LogScore} &= \frac{1}{R} \sum_{r = 1}^R \ell_r
% }

% \noindent We compare the LogScore values among all candidate models.
% Since the LogScore is the average of negative loglikelihoods, a lower
% LogScore indicates a better performing model.

\hypertarget{model-fitting}{%
\subsection{Model fitting}\label{model-fitting}}

Models are fit using {\tt R} \citep{team2013r}. The first stage of the 
area-level models are fit using the \texttt{survey} package 
\citep{lumley2004analysis}.
% Appropriately survey-weighted direct estimates of the area-level means
% are calculated using the \texttt{svymean()} function, and design-based
% covariance matrices were extracted via the \texttt{vcov()} function
% applied to the output of \texttt{svymean()}. 
The second-stage of the area-level models, as well as the unit-level 
models, are fit using Integrated Nested Laplace Approximation (INLA) 
\citep{rue2009approximate} 
as implemented in the \texttt{INLA} package. For the second stage of the
area-level models, since the \texttt{INLA} package does not provide a
bivariate likelihood, we must instead specify a univariate Gaussian
likelihood and add bivariate Gaussian IID random effects for each
region. These have fixed covariances equal to \(\hat{\bm{V}}^{des}_{r}\).

The models in this paper are easy to specify and fast to compute using
personal computers, generally taking less than 10 minutes. 
All code used in this paper is available at
\url{https://github.com/aeschuma/multivariate-sae}.

\hypertarget{simulation-studies}{%
\section{Simulation studies}\label{simulation-studies}}

We perform simulation studies to show the validity of our modeling
procedure and explore how misspecification of the spatial models---particularly 
the shared components---affects the results. We will summarize the simulation
studies here, while full descriptions and results can be found in Appendix B.

The simulations for area- and unit-level models are structured similarly. First, data 
similar to the working example of HAZ and WAZ scores in the 2014 KDHS are generated 
for multiple scenarios, each with their own data generating mechanism. The unit-level 
simulation study has seven scenarios for which we generate individual level data 
using parameter values set equal to those estimated from similar models fit to the 
2014 KDHS data. The area-level simulation study has nine scenarios; rather than
generating individual-level data, we simulate first-stage direct estimates and the 
associated variance estimates from their sampling distributions. 
Next, we fit seven second stage smoothing models for the area-level simulations and 
four models for the unit-level simulations that cover a range of univariate and bivariate 
models with and without shared components of various random effect structures. Finally, 
we compare bias, variance, and mean squared error (MSE), along with the width and 
coverage of 95\% credible intervals.

\subsection{Area-level simulation}

Models used in the area-level simulation include univariate and bivariate specifications
that vary in their spatial dependence and presence of shared components. 
We present the candidate models in Table 1.

\begin{table}[hbt!]
\caption{\label{tab:p2-candidate-models}Candidate models to estimate HAZ and WAZ from the 2014 KDHS. Model components are color-coded based on the corresponding descriptor in the model name. Model numbers are color-coded to be consistent with figures that show results from simulation studies.}
\centering
\begin{tabular}[t]{ll>{\raggedright\arraybackslash}p{7em}l}
\toprule
Model  & Model & Stage 1  & Stage 2 \\
Number &Name & Variance &Linear Predictor\\ 
\midrule
\color{gg1}O & \color{blue}Bivariate \color{black}Direct & $\color{blue}\begin{bmatrix} \hat{V}^{des}_{r11} & \hat{V}^{des}_{r12} \\ \hat{V}^{des}_{r21} & \hat{V}^{des}_{r22} \end{bmatrix}$ & \quad---\\
\color{gg2}I & \color{red}Univariate \color{magenta} IID & $\color{red}\begin{bmatrix} \hat{V}^{des}_{r11} & 0 \\ 0 & \hat{V}^{des}_{r22} \end{bmatrix}$ & $\begin{array}{l}\beta_1 + \color{magenta}v_{r1} \\ \beta_2 + \color{magenta}v_{r2} \end{array}$\\
\color{gg3}II & \color{red}Univariate \color{green} BYM & $\color{red}\begin{bmatrix} \hat{V}^{des}_{r11} & 0 \\ 0 & \hat{V}^{des}_{r22} \end{bmatrix}$ & $\begin{array}{l} \beta_1 + \color{green}v_{r1} + u_{r1} \\ \beta_2 + \color{green}v_{r2} + u_{r2}\end{array}$\\
\color{gg4}III & \color{blue}Bivariate \color{black} non-shared \color{magenta}IID & $\color{blue}\begin{bmatrix} \hat{V}^{des}_{r11} & \hat{V}^{des}_{r12} \\ \hat{V}^{des}_{r21} & \hat{V}^{des}_{r22} \end{bmatrix}$ & $\begin{array}{l}\beta_1 + \color{magenta}v_{r1} \\ \beta_2 + \color{magenta}v_{r2}\end{array}$\\
\color{gg5}IV & \color{blue}Bivariate \color{black} non-shared \color{green}BYM & $\color{blue}\begin{bmatrix} \hat{V}^{des}_{r11} & \hat{V}^{des}_{r12} \\ \hat{V}^{des}_{r21} & \hat{V}^{des}_{r22} \end{bmatrix}$ & $\begin{array}{l} \beta_1 + \color{green}v_{r1} + u_{r1} \\ \beta_2 + \color{green}v_{r2} + u_{r2}\end{array}$\\
\addlinespace
\color{gg6}V & \color{blue}Bivariate \color{cyan}Shared \color{magenta}IID & $\color{blue}\begin{bmatrix} \hat{V}^{des}_{r11} & \hat{V}^{des}_{r12} \\ \hat{V}^{des}_{r21} & \hat{V}^{des}_{r22} \end{bmatrix}$ & $\begin{array}{l} \beta_1 + \color{magenta}v_{r1}\color{black} + \color{cyan}\lambda (v_{2r}) \\ \beta_2 + \color{magenta}v_{r2}\end{array}$\\
\color{gg7}VI & \color{blue}Bivariate \color{cyan}Shared \color{green}BYM & $\color{blue}\begin{bmatrix} \hat{V}^{des}_{r11} & \hat{V}^{des}_{r12} \\ \hat{V}^{des}_{r21} & \hat{V}^{des}_{r22} \end{bmatrix}$ & 
$\begin{array}{l} \beta_1 + \color{green}v_{r1} + u_{r1}\color{black} + \color{cyan}\lambda (v_{r2} + u_{r2}) \\ \beta_2 + \color{green}v_{r2} + u_{r2}
\end{array}
$\\
\bottomrule
\end{tabular}
\end{table}

\begin{figure}

{\centering \includegraphics[width=0.95\linewidth]{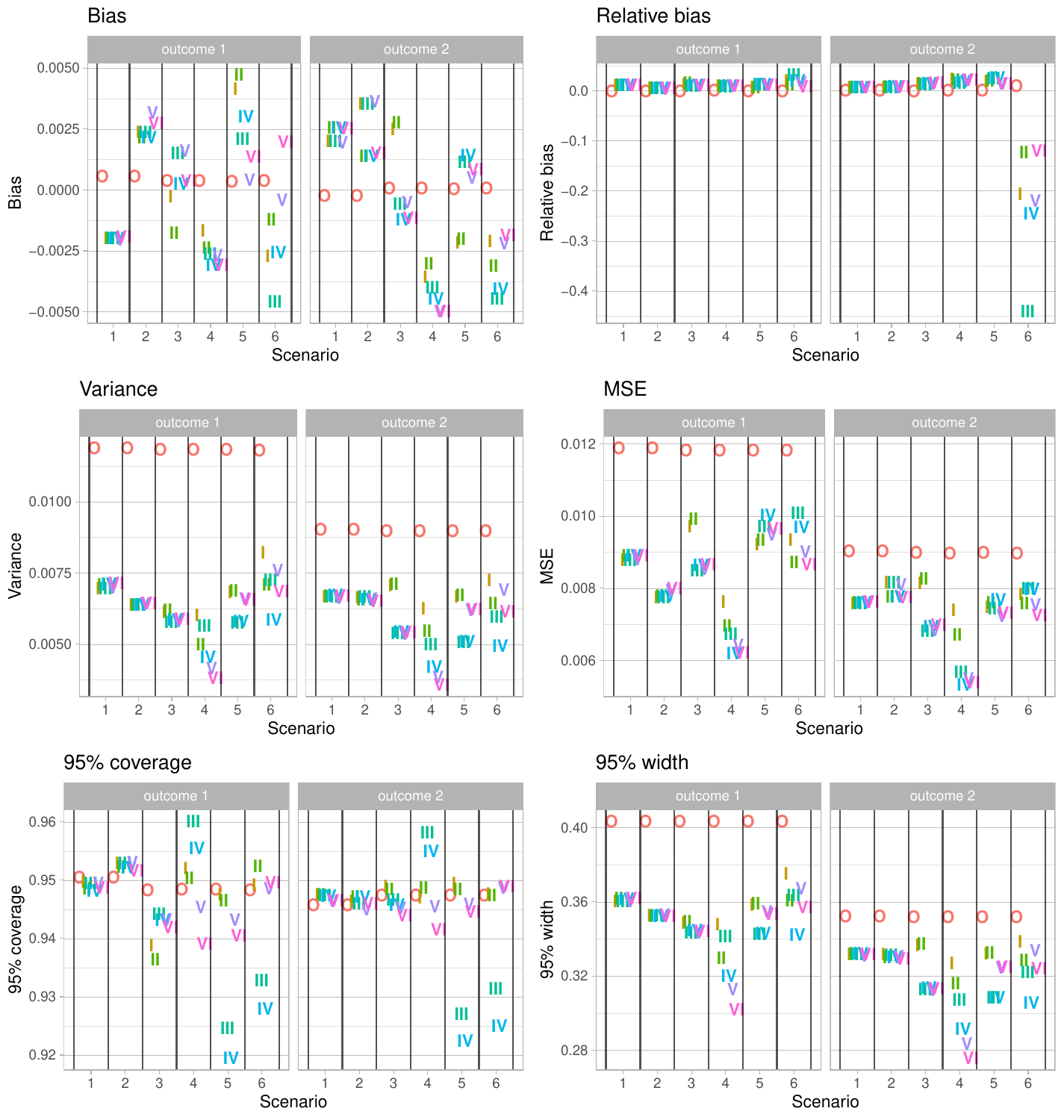} 

}

\caption{Bias, absoulte bias, variance, MSE, 95\% interval coverage, and 95\% interval width for all models across simulation scenarios 1--6. Scenario descriptions: 1 - univariate IID; 2 - univariate BYM; 3 - bivariate non-shared IID; 4 - bivariate non-shared BYM; 5 - bivariate shared IID; 6 - bivariate shared BYM. Models: \color{gg1}O - direct estimates; \color{gg2}I - univariate IID; \color{gg3}II - univariate BYM; \color{gg4}III - bivariate non-shared IID; \color{gg5}IV - bivariate non-shared BYM; \color{gg6}V - bivariate shared IID; \color{gg7}VI - bivariate shared BYM.}\label{fig:p2-sim-res-graph-1thru6}
\end{figure}

\begin{figure}

{\centering \includegraphics[width=0.85\linewidth]{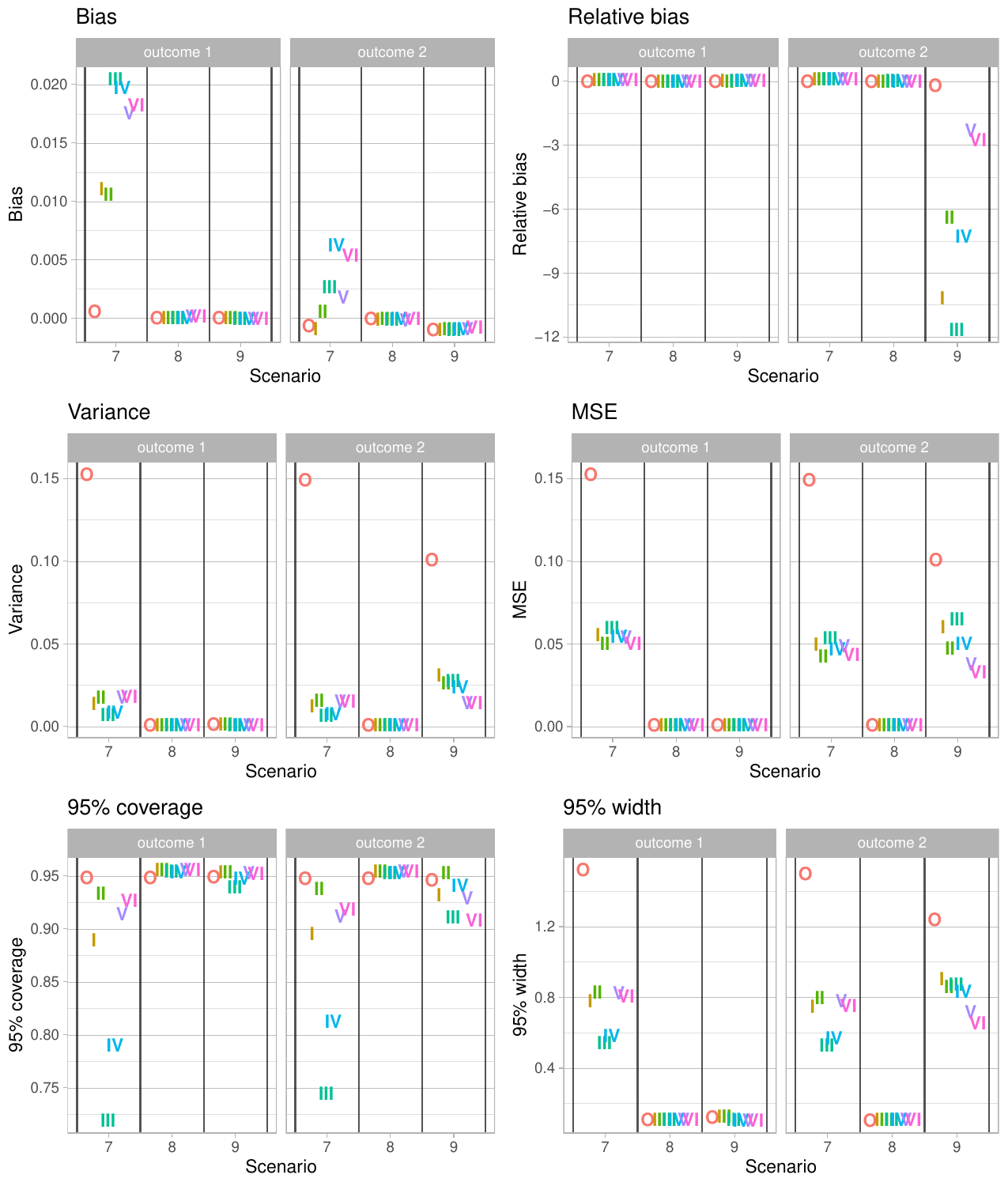} 

}

\caption{Bias, absoulte bias, variance, MSE, 95\% interval coverage, and 95\% interval width for all models across simulation scenarios 7--9. Scenario descriptions: 7 - bivariate shared BYM, large first stage variances; 8 - bivariate shared BYM, small first stage variances; 9 - bivariate shared BYM, correlated one large and one small first stage variance. Models: \color{gg1}O - direct estimates; \color{gg2}I - univariate IID; \color{gg3}II - univariate BYM; \color{gg4}III - bivariate non-shared IID; \color{gg5}IV - bivariate non-shared BYM; \color{gg6}V - bivariate shared IID; \color{gg7}VI - bivariate shared BYM.}\label{fig:p2-sim-res-graph-7thru9}
\end{figure}

Data for scenarios 1--4 are generated without shared components, and cover 
univariate and bivariate models with and without spatial correlation. 
Simulation results are shown in Figure 3.
In these scenarios, we see little bias across candidate models.
The direct estimates are the least biased, but have the
highest variance and MSE, as well as substantially wider intervals than the rest. 
In scenarios 1--3, all smoothing models perform comparably in terms of
variance, MSE, coverage, and width. In scenario 4, in which data are
generated via a bivariate non-shared BYM mechanism, the bivariate
non-shared BYM and bivariate shared BYM models have the lowest MSE and
similar close to nominal coverage, but the shared model actually has the
narrowest intervals which indicates it performing better than the
correctly specified model. 

Data for scenarios 5 and 6 are generated with shared components. 
Simulation results are also shown in Figure 3. 
We see similar results in terms of bias (relative bias for outcome 2 in scenario 6 
is driven by a small number of near-zero true values), and again the
direct estimates have the highest variance and MSE. However, the
bivariate non-shared models have undercoverage while the other models all
perform well. Additionally, the bivariate shared BYM model performs best
when it is the correct model, and the bivariate shared IID model
performs best when it is the correct model, but these models are quite
similar in both scenarios. This stresses the importance of accounting
for a shared component when there indeed is one and you are fitting a
bivariate model. From these results, we can conclude that with data the
size of the 2014 KDHS with outcomes similar to HAZ and WAZ, the choice
of second stage model is fairly unimportant as long as a shared
component is specified when one truly exists---but some sort of
second-stage smoothing is beneficial compared to the direct estimates in
order to narrow the uncertainty intervals without adding bias and still
preserving close to nominal coverage.

In scenarios 7 and 8, we generate data with shared spatial components---the
first stage variability is much higher in scenario 7 and much lower in scenario 8. 
Simulation results are shown in Figure 4.
These mimic having either less or more data than the 2014 KDHS and/or having outcomes 
with lower or higher variability (e.g., measured with high imprecision). 
For scenario 8, we draw similar conclusions as scenarios 1--6. For scenario
7, the direct estimates perform similarly to scenarios 1--6, and the bivariate shared 
BYM model has the least variance, MSE, and interval width, although the other 
models are close. We again see that the bivariate non-shared models have the 
narrowest uncertainty intervals, but these are too narrow as they also demonstrate 
substantial undercoverage. The univariate BYM and bivariate shared BYM models 
are the best performing, with slightly better coverage for the univariate
BYM but narrower intervals for the bivariate shared BYM model. The takeaway here 
is that the most important modeling choice is the spatial component---if data truly
have a shared component, using a shared model is important to prevent undercoverage.

In scenario 9, data is generated with one precise outcome and one outcome 
with high variance. Simulation results are shown in Figure 4.
We draw similar conclusions as scenarios 1--6.
This means that for scenario 9, we do not see that a clearly better
performance of the shared component model, because the bulk of the work
accounting for the correlation between the outcomes is done in the first
stage model which is the same in all bivariate models. However, the
univariate models do not perform too poorly, which means that the
utility of these models isn't too important in this particular scenario
with one large and one small variance for two correlated outcomes.
Further simulation scenarios may be able to tease out specific scenarios
where a bivariate shared modeling approach has better performance
compared to incorrectly specified models.

Appendix B contains tables of simulation results for all seven models in all nine scenarios.

\subsection{Unit-level simulation}

Models used in the unit-level simulations again vary in their spatial dependence
and presence of shared components. The four candidate models are IID non-shared, spatial non-shared, 
IID shared, and spatial shared, with spatial effects parameterized as 
BYM random effects. Appendix B provides detailed specifications of these models.

\begin{figure}

{\centering \includegraphics[width=0.85\linewidth]{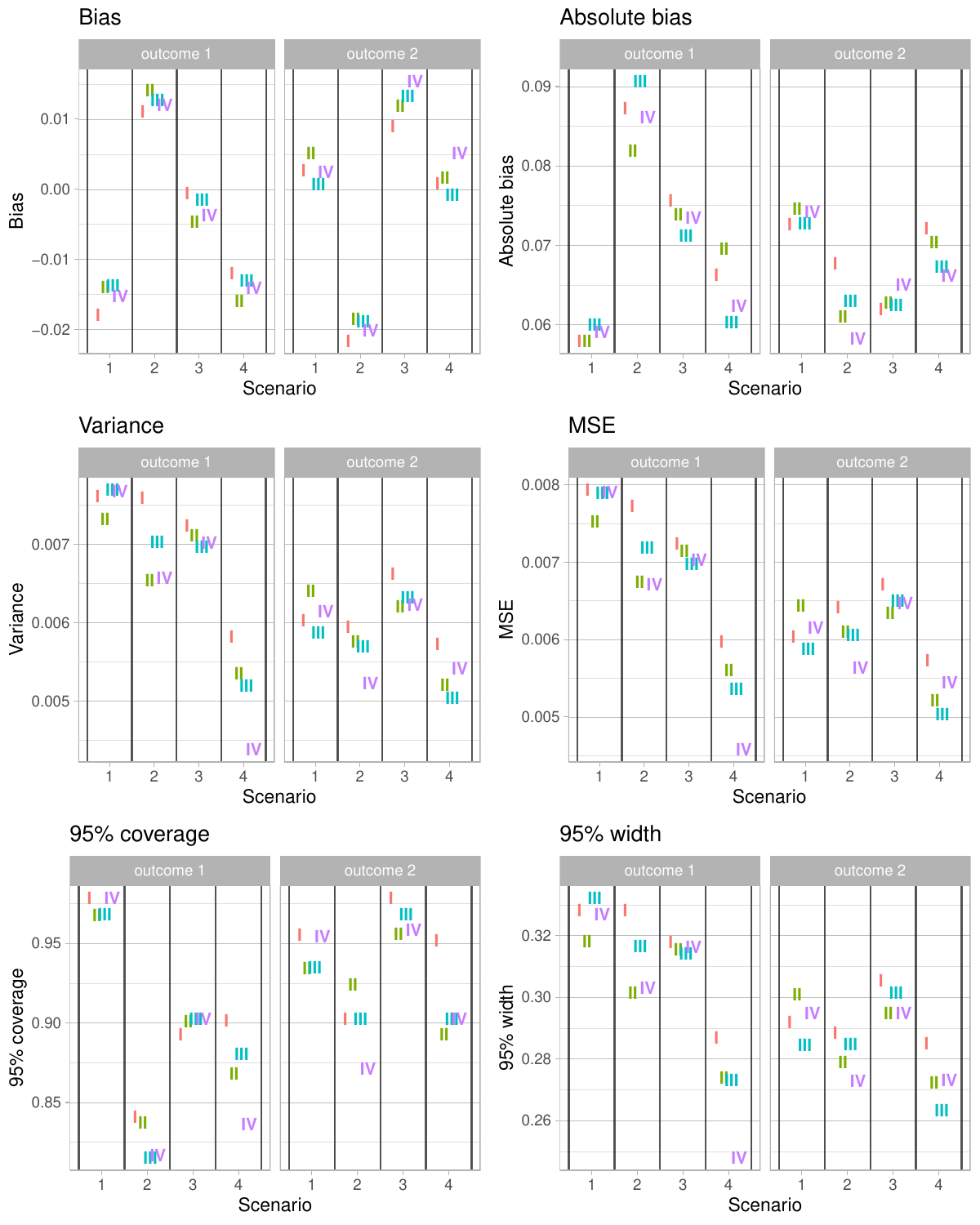} 

}

\caption{Bias, absoulte bias, variance, MSE, 95\% interval coverage, and 95\% interval width for all models across simulation scenarios 1--4. Scenario descriptions: 1 - IID non-shared; 2 - BYM non-shared; 3 - IID shared; 4 - BYM shared. Model descriptions: \color{gg1}I - IID non-shared; \color{gg3}II - BYM non-shared; \color{gg5}III - IID shared; \color{gg6}IV - BYM shared.}\label{fig:p3-sim-res-graph-1thru4}
\end{figure}

\begin{figure}

{\centering \includegraphics[width=0.85\linewidth]{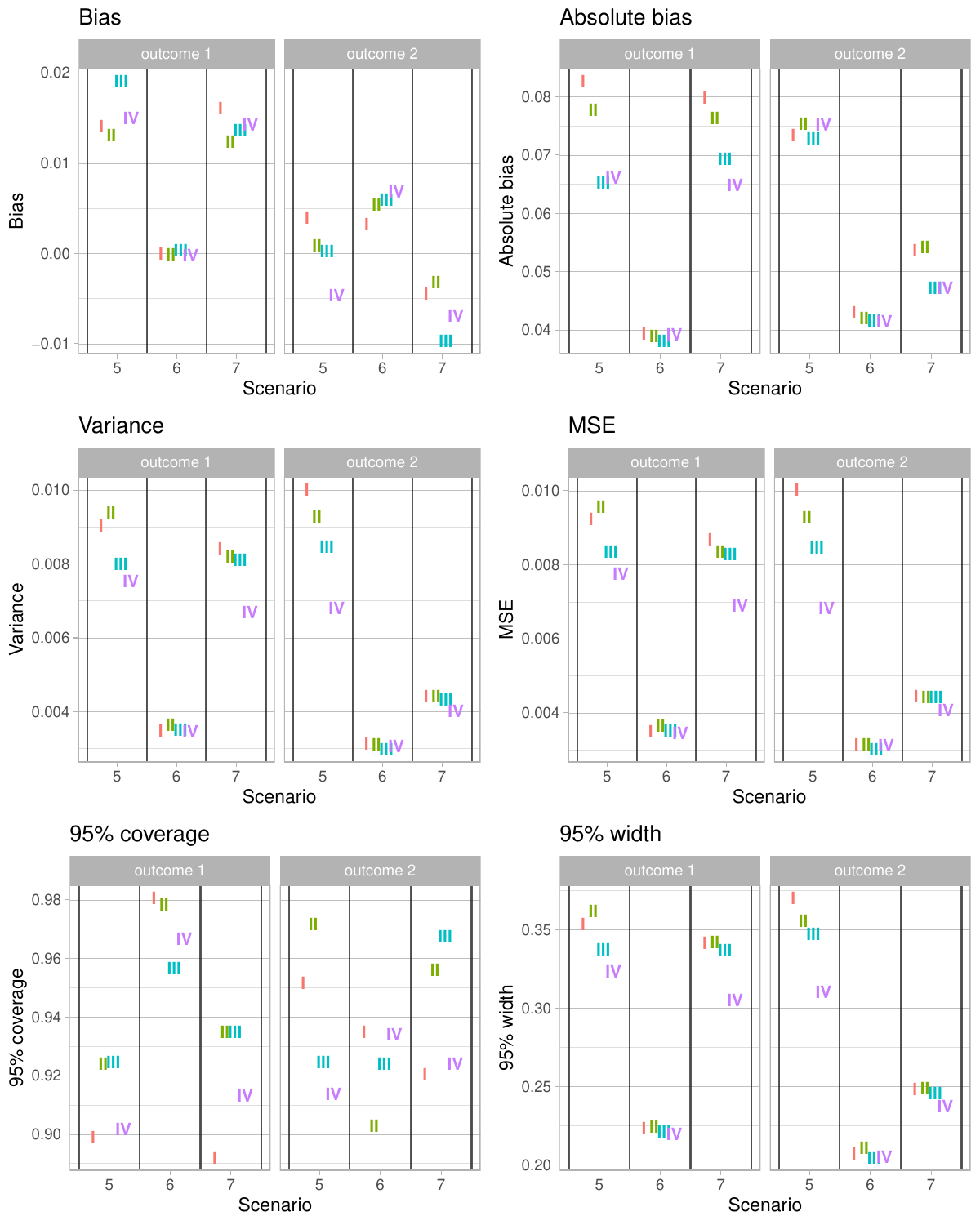} 

}

\caption{Bias, absoulte bias, variance, MSE, 95\% interval coverage, and 95\% interval width for all models across simulation scenarios 5--7. Scenario descriptions: 5 - BYM shared, larger variances; 6 - BYM shared, smaller variances; 7 - BYM shared, one precise and one imprecise outcome. Model descriptions: \color{gg1}I: IID non-shared; \color{gg3}II: BYM non-shared; \color{gg5}III: IID shared; \color{gg6}IV: BYM shared.}\label{fig:p3-sim-res-graph-5thru7}
\end{figure}

Scenarios 1--4 generate data as follows: (1) IID non-shared, (2) spatial non-shared, 
(3) IID shared, and (4) spatial shared. Simulation results are shown in Figure 5. 
We see little difference in bias among the 
models in these four scenarios. In scenarios 1--3, the variance,
MSE, coverage, and width are actually quite comparable, with no pattern
emerging of the correctly specified model outperforming the rest. In 
scenario 4, the correctly specified model does have the lowest MSE, which
is driven by having the lowest variance, but this leads to
the narrowest intervals and lowest coverage. This tradeoff
exposes a tricky decision when choosing a model, as the best
fitting model may provide estimates with inappropriately low variance leading to poor coverage. 

Scenarios 5--7 generate data with shared spatial components. Simulation results 
are shown in Figure 6. For scenario 6, 
with low variance of the Gaussian likelihood, all models perform similarly 
as there is little shrinkage. Scenario 5, with large Gaussian variance, and
scenario 7, with one larger and one smaller Gaussian variance,
demonstrate similar issues as seen in scenario 4. However,
scenario 7 also shows a markedly lower absolute bias in the outcome with
the larger variance, which indicates the utility of a shared component
to borrow information from a precise outcome to aid estimation of a less
precise outcome.

Appendix B contains tables of simulation results from all four models in all seven scenarios.

\hypertarget{modeling-examples}{%
\section{Modeling examples}\label{modeling-examples}}

In this section, we model HAZ and WAZ scores from the 2014 KDHS using
both area- and unit-level models as a working example for bivariate
continuous outcomes. Separately for area- and unit-level models, 
we present the modeling specifications for all candidate models,
compare them using the LogScore, and present detailed results for 
the best performing model. In Appendix A, we model contraceptive use 
from the 2014 KDHS data as a working example for multinomial outcomes.

\hypertarget{area-level-modeling-of-haz-and-waz}{%
\subsection{Area-level modeling of HAZ and
WAZ}\label{area-level-modeling-of-haz-and-waz}}

\hypertarget{area-level-model-specifications}{%
\subsubsection{Area-level model
specifications}\label{area-level-model-specifications}}

The candidate models in this section will all use the same first stage model 
but will have different second stage Bayesian smoothing models. The first stage 
model will be the bivariate case of the model in Section
3.2. With \(y_{ir1}\) and \(y_{ir2}\) the HAZ and WAZ
scores, respectively, for individual \(i\) in region \(r\), our first
stage model estimates are,
\eq{
    \hat{\bar{y}}_{r1} &= \frac{\sum_{i=1}^{n_r} w^{*}_{ri} y_{ri1}}{\sum_{i=1}^{n_r} w^{*}_{ri}} \\
    \hat{\bar{y}}_{r2} &= \frac{\sum_{i=1}^{n_r} w^{*}_{ri} y_{ri2}}{\sum_{i=1}^{n_r} w^{*}_{ri}}
}
\noindent with asymptotic normal likelihood
\eqnum{
    \hat{\bar{\bm{y}}}_r \mid  \bm{\mu}_r, \hat{\bm{V}}^{des}_{r} &\sim N_2\left(\bm{\mu}_r, \hat{\bm{V}}^{des}_{r} \right) \label{p2-eq-likelihood-2d}
}
\noindent where the estimated covariance matrix is denoted
\eq{
    \hat{\bm{V}}^{des}_{r} &= \begin{bmatrix} \hat{V}^{des}_{r11} & \hat{V}^{des}_{r12} \\ \hat{V}^{des}_{r21} & \hat{V}^{des}_{r22} \end{bmatrix}.
}

For the second stage, we use the same seven candidate models 
as the simulation in Section 4.1, which are described in Table 1. 
As we are using Bayesian inference, we must specify prior distributions.
This example has a large amount of data, so we use relatively
noninformative priors. For fixed effects, we use improper flat
priors. For IID random effects in IID-only models (I, III, and V),
we use penalized complexity (PC) priors \citep{simpson2017penalising} on
the standard deviations such that there is a 1\% probability that the standard deviation is greater
than 1. For BYM random effects, we use PC priors on the total variance
of the IID and ICAR random effects such that there is a 1\%
probability that the total standard deviation is greater than 1. We use a Beta(1, 1)
prior on each \(\rho_c\) as recommended in \citet{riebler2016intuitive}.

\hypertarget{area-level-modeling-results}{%
\subsubsection{Area-level modeling
results}\label{area-level-modeling-results}}

The bivariate correlations estimated from the first
stage model range from 0.34 to 0.95,
indicating a fairly strong relationship between HAZ and WAZ at the
cluster level (the shared component in the second stage
models will help capture the correlation between HAZ and WAZ at the 
regional level, i.e., due to unobserved covariates). This range of correlations
demonstrates the potential for improved estimates due to multivariate
modeling. We also note that the correlations are higher in regions with
more rural populations compared to urban populations, which lends some 
credence to unobserved covariates shared between HAZ and WAZ.

In Appendix B, we compare the posterior medians and standard deviations 
of the latent mean HAZ and WAZ scores among all models. The posterior median
estimates across the IID models generally agree, and the same is true
for BYM models. Comparing BYM models to IID models, the spatial smoothing 
causes the estimates to shrink toward the overall mean. Since there is an 
abundance of data, however, this shrinkage is less pronounced than would 
occur when analyzing data from a smaller survey. The non-shared BYM and
non-shared IID models have the smallest posterior standard deviations,
which lead to narrower uncertainty intervals that will be anticonservative 
in scenarios with a true shared component as discussed in Section 4.2
(for example, if there are unobserved covariates shared between the
outcomes).

Table 2 compares the LogScore (Section 3.4) for each of the seven
candidate models. The LogScore is akin to the
average of the posterior predictive negative log likelihood calculated 
via leave-one-out cross validation, and a
lower score means a better performing model. The bivariate shared BYM 
model has the lowest LogScore, likely indicating the presence of shared unobserved 
factors that influence both HAZ and WAZ. Thus, we present final results 
from this model. As a sensitivity analysis, we fit this model with a 
shared HAZ component, rather than a shared WAZ component, and estimates 
were nearly identical.

We summarize the posterior distributions of estimated parameters in Table 3, 
with parameters defined as in Section 3.
Figures 7 and 8 show maps of the posterior medians and 80\% credible
intervals, respectively, for HAZ and WAZ scores. We see a clearer spatial 
gradient for WAZ (lower in the north and higher in the south). This is 
reflected by \(\rho_2\) being greater than \(\rho_1\), indicating that WAZ 
has a comparatively higher proportion of the total random effect variance 
that is spatial. For WAZ, these findings are consistent with 
the general knowledge that access to and quality of health care and proper nutrients
tend to be worse in the northern regions \citep{ilinca2019socio}.
Furthermore, while areas of high uncertainty in HAZ generally
correspond to high uncertainty in WAZ, some regions do not exhibit 
this pattern.

We plot maps of the IID and ICAR random effects for both HAZ and WAZ in
Appendix B. The differences in the spatial gradients of the HAZ ICAR and
the shared ICAR components is particularly striking. This is because the
shared ICAR component captures the spatial variation in WAZ, while
the HAZ ICAR component captures the residual spatial variation beyond
that which is shared with WAZ. The spatial correlation in WAZ is stronger
than HAZ, resulting in opposite spatial gradients. We provide more
discussion of this phenomenon when examining the unit-level model for
these data in Section 5.2.

\begin{table}

\caption{\label{tab:p2-mod-comp-metrics}LogScore of area- and unit-level candidate models for estimating HAZ and WAZ from the 2014 KDHS data. Bold indicates the best performing model}
\centering
\begin{tabular}[t]{lrr}
\toprule
Model & Area-level LogScore & Unit-level LogScore\\
\midrule
Bivariate non-shared IID & 0.38 & 0.21 \\
Univariate IID & 0.27 & -- \\
Bivariate shared IID & 0.14 & 0.33 \\
Univariate BYM & -0.16 & -- \\
Bivariate non-shared BYM & -0.23 & -0.35 \\
\textbf{Bivariate shared BYM} & \textbf{-0.47} & \textbf{-0.45} \\
\bottomrule
\end{tabular}
\end{table}

\begin{table}

\caption{\label{tab:p3-posterior-ests}Posterior medians (Est) and 95\% credible intervals (CI) for fixed effects and hyperparameters from the area- and unit-level BYM shared models estimating HAZ and WAZ from the 2014 KDHS data. Subscript 1 refers to HAZ and subscript 2 refers to WAZ.}
\centering
\begin{tabular}[t]{lllll}
\toprule
Parameter & \multicolumn{2}{l}{Area-level} & \multicolumn{2}{l}{Unit-level} \\
 & Est & 95\% CI & Est & 95\% CI\\
\midrule
$\beta_1 \text{ (HAZ) }$ & -0.98 & (-1.03, -0.94) & -0.77 & (-0.83, -0.71)\\
$\beta_2 \text{ (WAZ) }$ & -0.87 & (-0.91, -0.83) & -0.66 & (-0.71, -0.61)\\
$\sigma_1$ & 0.19 & (0.15, 0.24) & 0.13 & (0.1, 0.17)\\
$\rho_1$ & 0.79 & (0.35, 0.96) & 0.71 & (0.25, 0.97)\\
$\sigma_2$ & 0.25 & (0.19, 0.31) & 0.19 & (0.16, 0.23)\\
$\rho_2$ & 0.92 & (0.57, 0.99) & 0.67 & (0.42, 0.88)\\
$\lambda$ & 0.75 & (0.47, 1.01) & 0.72 & (0.46, 1.01)\\
$\gamma_1$ & -- & -- & -0.29 & (-0.35, -0.23)\\
$\gamma_2$ & -- & -- & -0.29 & (-0.34, -0.23)\\
$\omega_1$ & -- & -- & 1.31 & (1.3, 1.32)\\
$\omega_2$ & -- & -- & 1.14 & (1.13, 1.15)\\
$\sigma_{\epsilon 1}$ & -- & -- & 0.36 & (0.34, 0.39)\\
$\sigma_{\epsilon 2}$ & -- & -- & 0.33 & (0.31, 0.35)\\
\bottomrule
\end{tabular}
\end{table}

\begin{figure}

{\centering \includegraphics[width=0.98\linewidth]{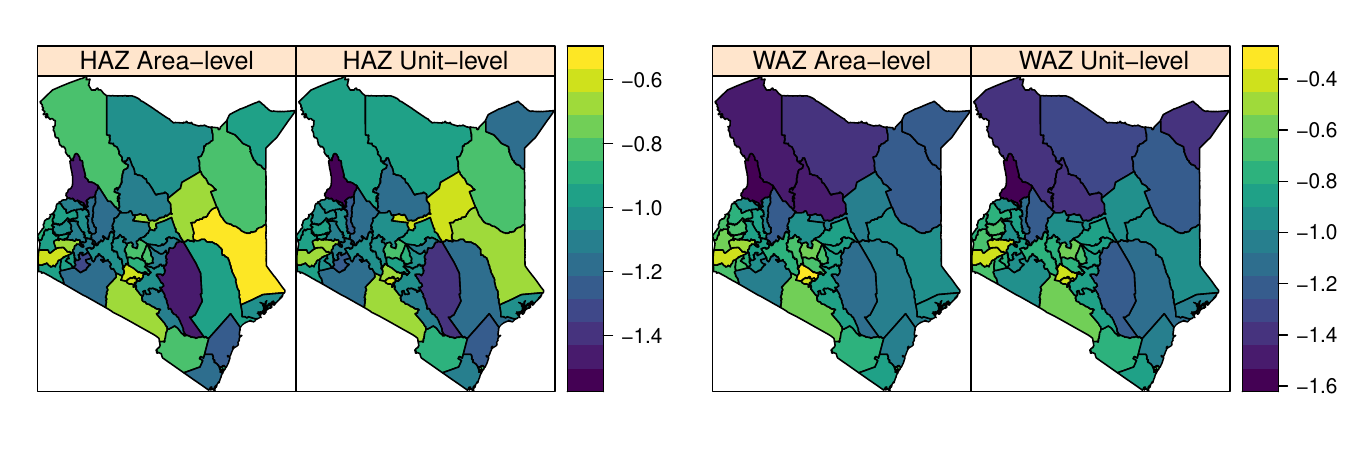} 

}

\caption{Estimated posterior medians from area-level Bivariate shared BYM model and unit-level BYM shared model fit to HAZ and WAZ from the 2014 KDHS.}\label{fig:concl-hazwaz-med-compare-map}
\end{figure}

\begin{figure}

{\centering \includegraphics[width=0.98\linewidth]{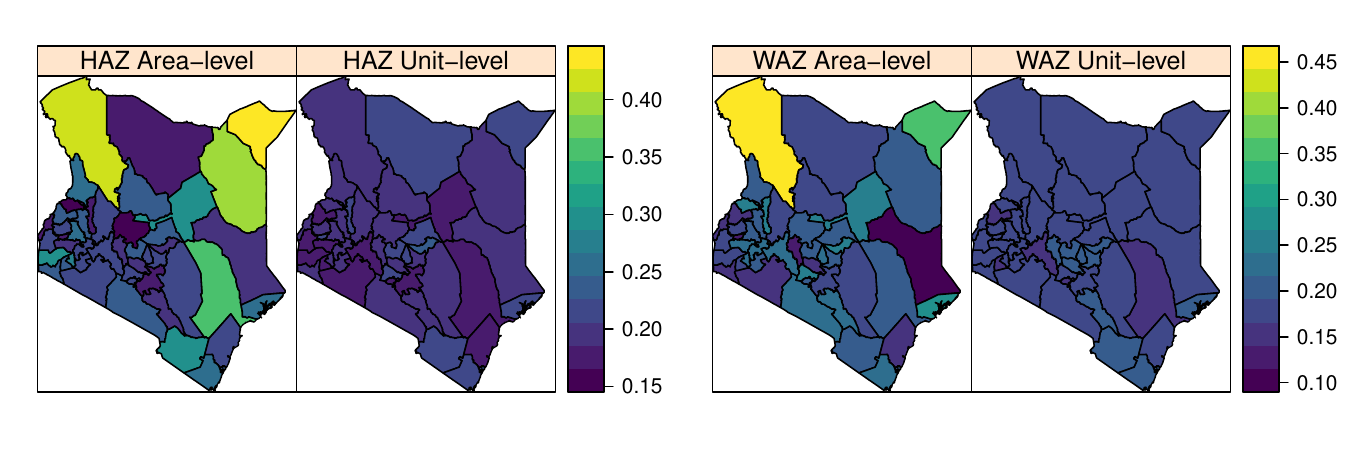} 

}

\caption{Estimated posterior 80\% credible interval width from area-level Bivariate shared BYM model and unit-level BYM shared model fit to HAZ and WAZ from the 2014 KDHS.}\label{fig:concl-hazwaz-width-compare-map}
\end{figure}

\hypertarget{unit-level-modeling-of-haz-and-waz}{%
\subsection{Unit-level modeling of HAZ and
WAZ}\label{unit-level-modeling-of-haz-and-waz}}

\hypertarget{unit-level-model-specifications}{%
\subsubsection{Unit-level model
specifications}\label{unit-level-model-specifications}}

We fit the same suite of unit-level models as the simulation study 
(Section 4.2, with models described in Appendix B), using similar 
prior distributions as the area-level second stage smoothing
models described in Section 5.1. To aggregate estimates to the area-level, we 
use the proportion of the 2014 under-5 population in each region that is 
rural. These were calculated from urban/rural classification and 
population density rasters created by WorldPop\citep{worldpop2022} in
conjunction with region-level urban population fractions from the 2014 KDHS 
using the thresholding method described in \citet{wu2021spatial}.

\hypertarget{unit-level-modeling-results}{%
\subsubsection{Unit-level modeling
results}\label{unit-level-modeling-results}}

In Appendix B, we compare the posterior medians and standard deviations
of the aggregated area-level HAZ and WAZ estimates among the four candidate models. 
All models show strong agreement, likely due to the large amount of data. 
The BYM shared models have the smallest posterior standard deviations. 
However, as discussed in Section 4.2, these may lead to uncertainty intervals 
with frequentist coverage below the nominal level.

We compare the LogScore for each model in Table 2. The shared BYM model
performs best, which is intuitive given the strong spatial dependence
between HAZ and WAZ, as well as likely shared unobserved factors that influence both.
Thus, we present final results from this model with the caveat that uncertainty intervals 
might be too narrow.

Summaries of posterior distributions are presented in Table 3, with 
parameters defined as in Section 3. Figures 7 and 8 show maps of the 
posterior medians and 80\% credible intervals, respectively, for
HAZ and WAZ scores. Similar to the area-level results, the WAZ estimates 
display a more consistent spatial gradient compared to HAZ. 

To explore further, we show maps of the total BYM random effects along 
with the standardized IID and ICAR components in Appendix B.
The north-south heterogeneity is evident in the shared ICAR random
effect, with additional hotspots of high WAZ in the south-center and
southwest which arise from the shared IID random effect. This is the
complete picture for the WAZ final estimates, but recall that the HAZ
final estimates include both the shared random effects and HAZ-specific
random effects. Thus, we can interpret the HAZ ICAR random effects as
the difference between the WAZ spatial gradient and the HAZ spatial 
gradient; similarly, the HAZ IID random effects constitute the non-structured 
differences. Since the HAZ random effects are generally in the opposite 
direction as the shared random effects, the north-south gradient for HAZ 
becomes less stark, especially in the northern regions. 

The spatial distribution seen here, which is also observed in other health 
indicators in Kenya \citep{mulatya2020disease, godwin2021space}, is commonly 
attributed to the rural-urban health divide \citep{joseph2020spatial}. However, 
the geographical heterogeneity estimated from the spatial random effects is
\emph{in addition to} the strong urban/rural effect estimated
in our model (\(\hat\gamma_1 = -0.29, \hat\gamma_2 = -0.29\)).
This indicates complex interactions of multiple factors influencing the spatial
distribution of HAZ and, more strongly, WAZ, beyond the urban-rural
gap. This is supported in recent literature \citep{fagbamigbe2020demystifying}.

\hypertarget{discussion}{%
\section{Discussion}\label{discussion}}

Jointly modeling multivariate outcomes is a useful endeavor to
produce more reliable estimates of key demographic and health indicators
in LMICs compared to estimates from separate univariate models. This
paper developed both area- and unit-level shared component SAE models
with discrete spatial effects to jointly model multiple outcomes. We
described how to select between models and provided examples by
producing estimates of HAZ and WAZ as well as contraceptive
use using data from the 2014 KDHS. These classes of models provide novel 
methods for estimating multivariate health and demographic outcomes from 
complex survey data that are affected by underlying unobserved covariates. 
Their utility was demonstrated through simulation studies and
modeling examples in which they performed best out of a suite of alternatives.

One aspect of the motivating data not considered was
measurement error. For HAZ and WAZ, measurement is notoriously difficult
\citep{ulijaszek1999anthropometric}. If the measurement error is not
systematically biased, then the uncertainty of our estimates will not
reflect the true uncertainty in the data, thus being anticonservative.
Any bias, however, will propagate through and affect the mean estimates.
For example, if a specific DHS interviewer that only worked in one
geographic region tended to under-weigh children, then WAZ in those
regions would be artificially lower, and our spatial estimates would be
less accurate. Future work is critical to assess the impact of
measurement error, and analyses should consider this when interpreting results. 
Adapting these models to account for measurement error may be a fruitful pursuit.

Our modeling examples use relatively uninformative PC priors. Due to the 
amount of data available, this is an adequate choice. Analyses with more 
informative prior distributions had negligible differences in estimates 
(results not shown). However, in other situations with less data or more 
variable outcomes, the choice of prior distribution may be important. Further
research can explore how different priors affect resulting estimates, 
especially priors on the shared component.

This paper restricted the class of spatial models used to those with BYM 
effects that use an ICAR model for spatial dependence, which are commonly 
used \citep{mercer2015space, wakefield2020small}. Other spatial models
could be considered, such as conditional autoregressive models or models with
Leroux spatial random effects \citep{leroux2000estimation}, though we 
suspect the form of the spatial effects may not be particularly 
impactful. Additionally, our work can be
extended to include spatiotemporal models. Spatiotemporal
Fay-Herriot modeling approaches have been applied in many analyses
\citep{marhuenda2013small, esteban2016area, rumiati2019spatio, li2019changes}.
A spatiotemporal extension of shared component models could be used in
these contexts, such as jointly modeling under-5 mortality from
different causes or jointly modeling HIV prevalence by contraceptive
use. Recent developments on spatiotemporal shared component models
\citep{paradinas2017spatio, mahaki2018joint, ahmadipanahmehrabadi2019bivariate, blangiardo2020advances}
can be explored to examine whether these approaches translate to the
context of estimating demographic and health outcomes from complex
survey data in LMICs.

This leads to a discussion of how to develop models in more complex scenarios. 
One particularly interesting case is modeling cause-specific mortality.
Here, the area-level first stage model becomes much more complicated. 
Rather than simply aggregating means appropriately with regard to the 
survey sampling design, we must account for the time-to-event nature of the
data, which is commonly done with discrete time survival analysis
techniques \citep{mercer2015space}. Our proposed modeling framework can be 
adapted accordingly, but this requires a large amount of data. In preliminary
work modeling cause-specific under-5 mortality using data from the
verbal autopsy module in the 2017 Bangladesh DHS \citep{bangladesh2020bangladesh}, 
there was not enough data to produce stable first stage estimates, let 
alone fit second stage models. Some LMICs collect data
from sources other than household surveys, notably sample vital registration
systems \citep{rao2010mortality, nkengasong2020improving}, which have been
used to estimate health indicators such as cause-specific child mortality 
\citep{he2017national, schumacher2022flexible}. Developing methods to use 
multiple types of data is an appealing avenue to research, and the 
methods covered in this paper will be a good starting point. 

% Another interesting route to explore is comparing shared component
% models with models that include covariates. Since the primary mechanism
% through which we motivated the shared component model is via shared but
% unobserved risk factors and other covariates, it would be interesting to
% see how models with covariates compare to shared component models
% without covariates. This exercise may provide further support for our
% class of shared component models as a whole, and it is straightforward
% to include area-level covariates.

\begin{figure}[ht]

{\centering \includegraphics[width=0.95\linewidth]{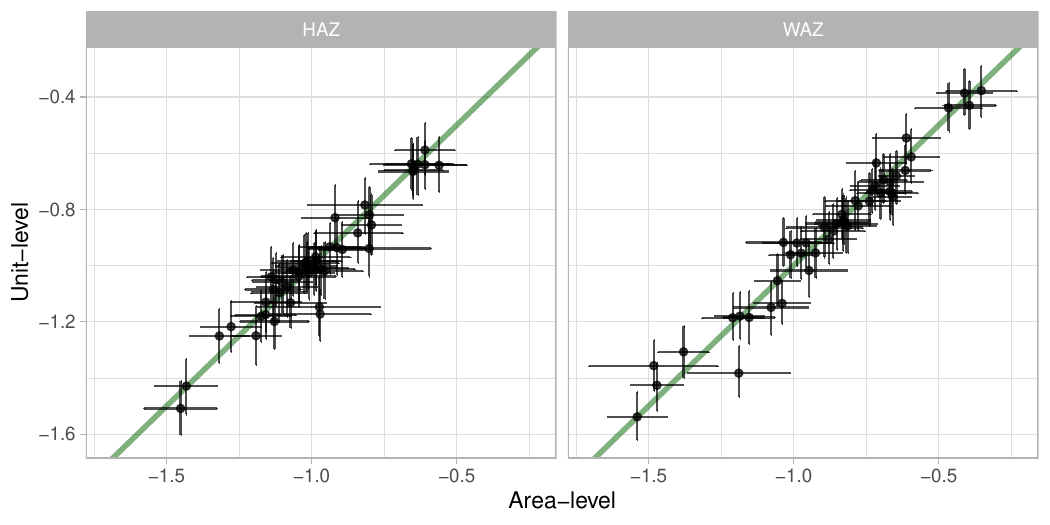} 

}

\caption{Estimated posterior medians and 80\% credible intervals from area-level Bivariate shared BYM models vs. unit-level BYM shared models fit to HAZ and WAZ from the 2014 KDHS.}\label{fig:concl-hazwaz-compare-scatter}
\end{figure}

Another important issue is comparing area- and unit-level models.
Figure 9 shows scatterplots of posterior medians with 80\% credible intervals 
for HAZ and WAZ that compare the final area- and unit-level models used in our working 
example. While the point estimates are quite similar, the unit-level model has 
narrower uncertainty intervals. These findings are emphasized in the maps
presented in Figures 7 and 8 and are consistent with the unit-level modeling
simulation, in which the uncertainty intervals for most scenarios exhibit coverage 
lower than the nominal level. This is a prime issue to consider when choosing 
between approaches. If uncertainty intervals for area-level models are reasonable 
to draw reliable conclusions, then area-level models are preferred.
However, when less data are available or data are noisier, area-level
models may exhibit too much posterior uncertainty. In this case,
unit-level models are desirable, although work is needed to understand the 
reasons for the undercoverage we have seen here.

% In the example of modeling contraceptive use data, we see that the
% posterior median estimates from the area- and unit-level models were similar for
% the two most prevalent categories, whereas the category with few
% observations saw larger differences. Some of this discrepancy may be
% due to the addition of ``phantom'' observations in the area-level
% models, as the regions where these observations were added
% had the largest differences. Hence, using unit-level models in
% situations where 0 counts are observed may lead to more consistent
% estimates in the smallest domains. However, the undercoverage of
% estimates remains an issue with unit-level models.

The novel approaches in this paper fill crucial methodological
gaps in the literature of modeling health and demographic outcomes in
LMICs. Future work remains to compare these approaches in more detail and 
in different scenarios, including investigating their utility as the number 
of outcomes increases. Extensions into spatiotemporal models are also an 
exciting next step due to the need for estimating time trends and
forecasting. Adaptations that facilitate estimation at 
finer spatial disaggregation will also be important.

\renewcommand\refname{References}

\bibliography{mybib.bib}

\end{document}